\documentclass[prx,twocolumn,amsmath,amssymb]{revtex4}

\usepackage{color}
\usepackage{bm}
\usepackage{dcolumn}% Align table columns on decimal point
\usepackage[dvips]{graphicx}

\usepackage{ulem}

\newcommand{\1}{\mbox{1}\hspace{-0.25em}\mbox{l}}

\newcommand{\tr}{{\rm tr}\,}
\begin{document}

\preprint{preprint}

\title{
Entanglement polarization for  the topological quadrupole phase
}

\author{Takahiro Fukui$^1$}
\author{Yasuhiro Hatsugai$^2$}
\affiliation{$^1$Department of Physics, Ibaraki University, Mito 310-8512, Japan}
\affiliation{$^2$Institute of Physics, University of Tsukuba, 1-1-1 Tennodai, Tsukuba, Ibaraki 305-8571, Japan}

\date{\today}

%%% abstract %%%
\begin{abstract}
We propose the entanglement dipole polarization to describe the topological quadrupole phase.
The quadrupole moment can be regarded as a pair of the dipole moment, in which the total dipole moment is canceled.
The entanglement polarization, we propose,  is useful to detect such a constituent dipole polarization. 
We first introduce partitions of sites in the unit cell and divide the system into two subsystems.
Then, introducing an
entanglement Hamiltonian  by tracing out one of the subsystems partly,
we compute the dipole polarization of the occupied states associated with the entanglement Hamiltonian, 
which is referred to as the entanglement polarization.
       Although the total dipole polarization is vanishing, those of the
       subsystems can be finite.
The entanglement dipole polarization is quantized by reflection symmetries.
We also introduce the entanglement polarization of the edge states,
which reveals that the edge states
themselves are gapped and topologically nontrivial.
Therefore, such edge states yield the zero energy edge states if the system has boundaries. 
This is the origin of the corner states. 
\end{abstract}

\pacs{
}

\maketitle

\section{Introduction}

Electric dipole polarization \cite{Vanderbilt:1993fk,King-Smith:1993aa}, known also as the Zak phase \cite{Zak:1989fk}
(Berry phase \cite{berry:1984aa}) ,
plays a key role in the description of topological phases of matter \cite{Hasan:2010fk,Qi:2011kx}.
One of the typical examples is the Su-Schrieffer-Heeger model in one dimension \cite{Su:1979aa,Su:1980aa}
in which the topological phase is labeled by the dipole polarization;
that is, the Berry phase, $\pi$ \cite{Ryu:2002fk,Hatsugai:2009}. 
Such a nontrivial polarization of the bulk ensures the zero-energy midgap edge states
for finite systems with boundaries \cite{Ryu:2002fk,Hatsugai:1993fk}.
In two dimensions, the nontrivial Chern number 
\cite{Thouless:1982uq,kohmoto:85} implies
  the existence of the topologically stable edge states \cite{Hatsugai:1993fk}
(bulk-edge correspondence).
The Chern number can also be interpreted by the change of the polarization (the Berry phase) in the Brillouin zone
\cite{Vanderbilt:1993fk,King-Smith:1993aa,Fu:2006aa}.
The charge polarization has been extended to time-reversal polarization \cite{Fu:2006aa,Qi:2008aa},
and to higher dimensions \cite{Qi:2008aa,Essin:2009aa,Essin:2010aa,Malashevich:2010aa,Marzari:2012aa}
which enables us to characterize the topological insulators in two and three dimensions.
Recently, a generalization of the dipole polarization,  i.e.,  quadrupole polarization and generically 
higher order multipole polarization has been proposed 
\cite{Benalcazar61},
and investigated in detail 
\cite{Benalcazar:2017aa,Liu:2017aa,Langbehn:2017aa,Song:2017aa,Hashimoto:2017aa,Schindler:2018aa,
1708.03647,Ezawa:2018aa,Ezawa:2018ab}.

In conventional topological insulators with spin-orbit couplings, the total dipole polarization vanishes, since 
each spin has just opposite polarization ensured by time-reversal symmetry.
Nevertheless, they can be topological if the polarization of each spin is nontrivial.
Such a time-reversal polarization is described by the celebrated Z$_2$ invariant \cite{Kane:2005aa,Fu:2006aa}.
Recently, an alternative method has been proposed to extract the polarization of each spin \cite{Fukui:2014qv,Fukui:2015fk}.
This method is based on the topological numbers of the entanglement Hamiltonian (eH).
It turns out that not only the topological numbers of the original Hamiltonian, but also those of eH
of subsystems are very useful to characterize the topological phases of complicated systems 
\cite{Araki:2016aa,Araki:2017aa}.

In this paper, we   demonstrate the use of topological numbers of  the eH
applying to the quadrupole phase in two dimensions.
This is based on the observation that the quadrupole moment is composed of two sets of the dipole moment. 
Therefore, if one of them is traced out, the other dipole moment should be revealed.
We argue that the dipole polarization of the ground state of the eH, which will be referred to as the entanglement polarization (eP),
is quite useful to characterize the quadrupole phase.

This paper is organized as follows:
In Sec. \ref{s:Model}, we review basic properties of the model introduced by Benalcazar {\it et. al.} \cite{Benalcazar:2017aa}.
In Sec. \ref{s:EntPol}, we introduce the eP and discuss the symmetry properties of the eP. 
Surprisingly, the exact eP can be obtained for the present model, implying the usefulness of the eP.
In Sec. \ref{s:eESP}, we focus our attention on the edge states of the model.
The model shows the gapped edge states, and eP for the edge states reveals that they are topologically nontrivial as one-dimensional (1D)   insulating states.
Therefore, if the boundary is introduced 
to these 1D topological insulating edge states, zero-dimensional (0D) edge states appear. 
It turns out such zero-energy edge states
are nothing but the corner states.
In Sec. \ref{s:GenBBH}, we apply eP to more generic model with next-nearest neighbor hopping terms.
In Sec. \ref{s:SumDis}, we give the summary and discussion.

\section{The model}
\label{s:Model}

We will refer to the two-dimensional variant of the SSH model introduced by Benalcazar {\it et. al.} \cite{Benalcazar:2017aa}
as the BBH model.
In this section, we define the notations of the BBH model, including a few review.
\begin{figure}[htb]
\begin{center}
\includegraphics[scale=0.4]{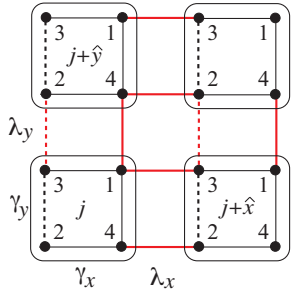}
\caption{2D BBH model on the square lattice. The dashed lines show the negative bonds associated with the $\pi$-flux.
}
\label{f:ssh}%-----------------------------------------------
\end{center}
\end{figure}

\subsection{The BBH model}

\begin{figure}[htb]
\begin{center}
\begin{tabular}{cc}
\includegraphics[scale=0.3]{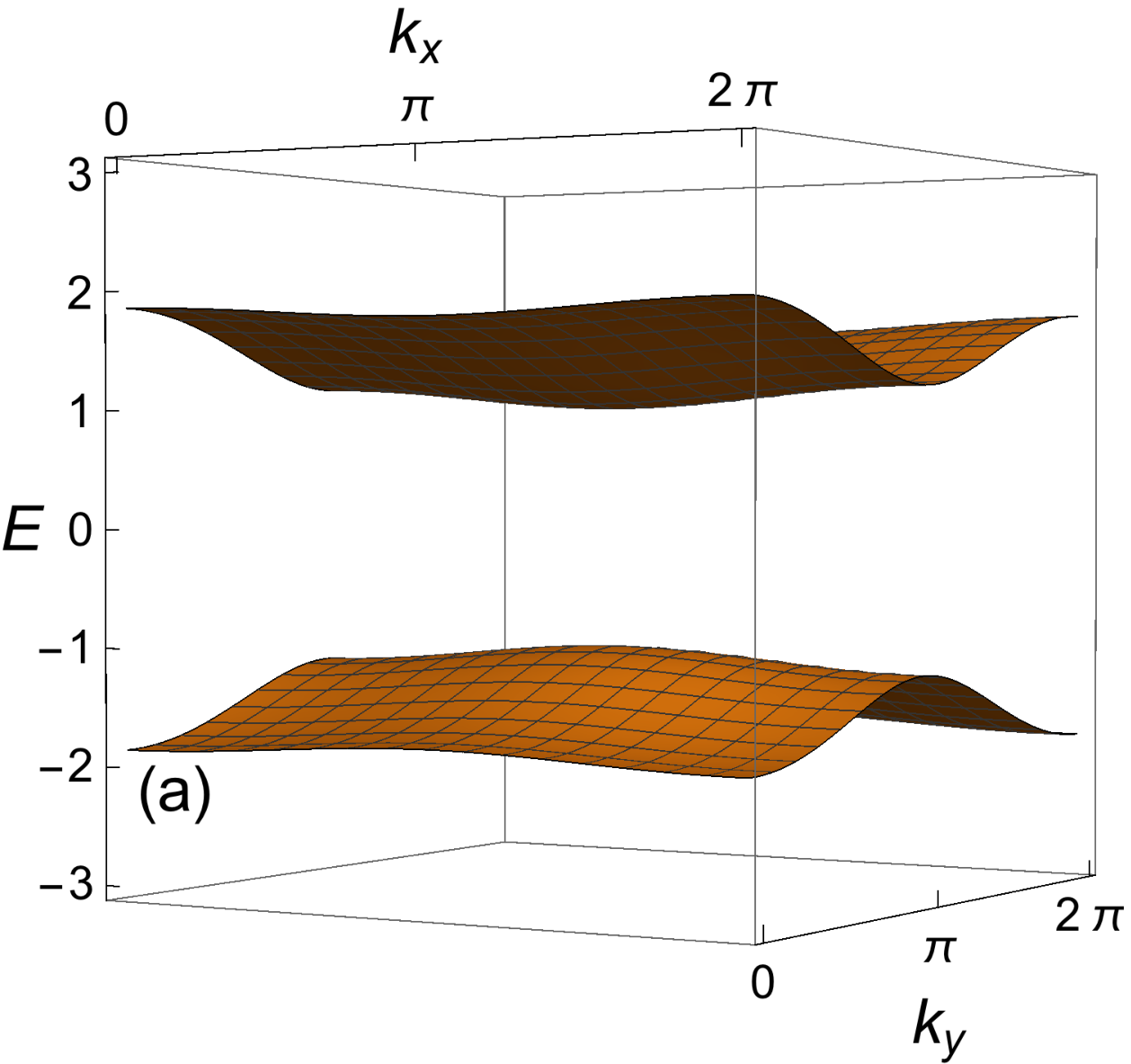}
&
\includegraphics[scale=0.3]{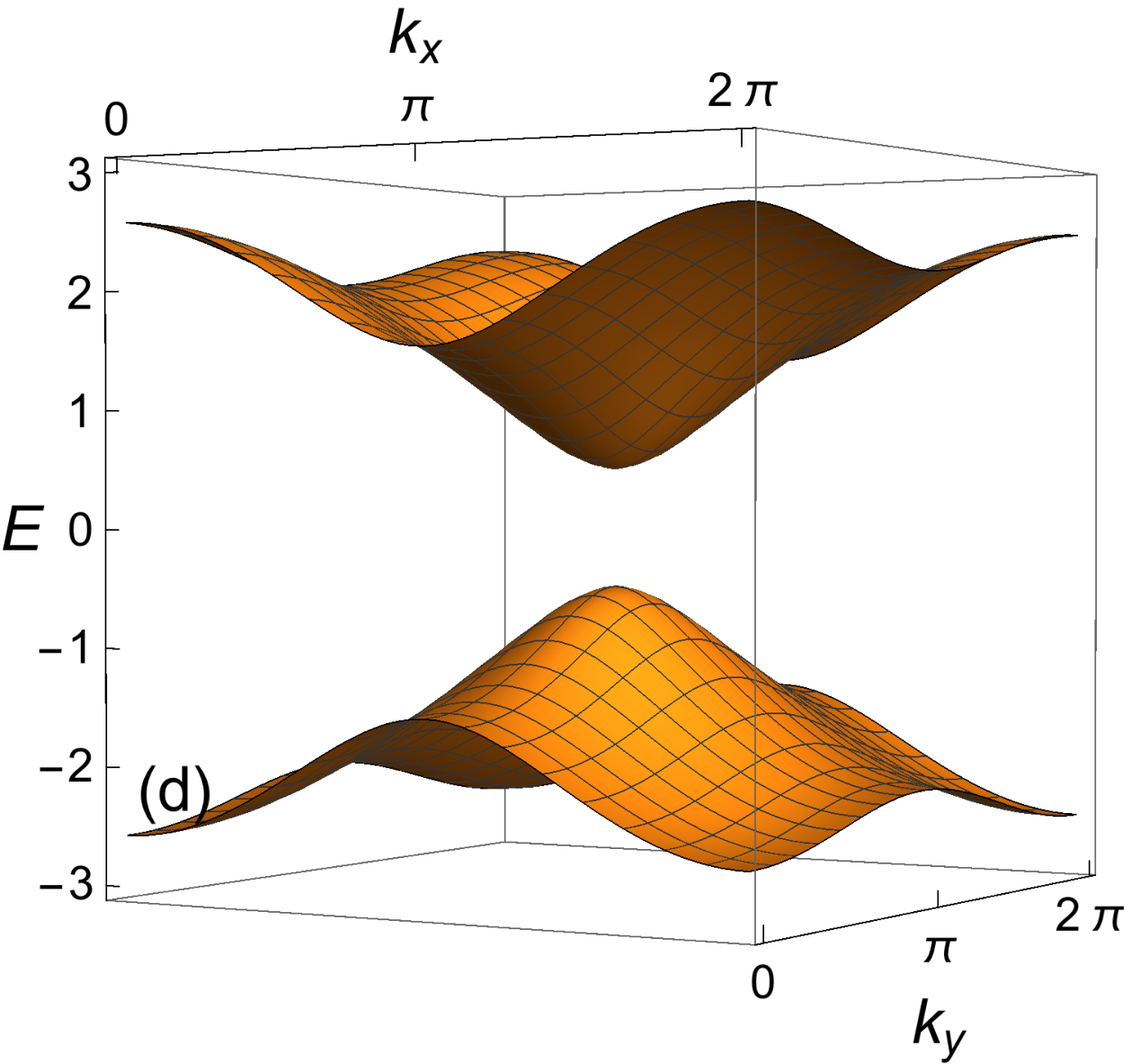}
\\
\includegraphics[scale=0.33]{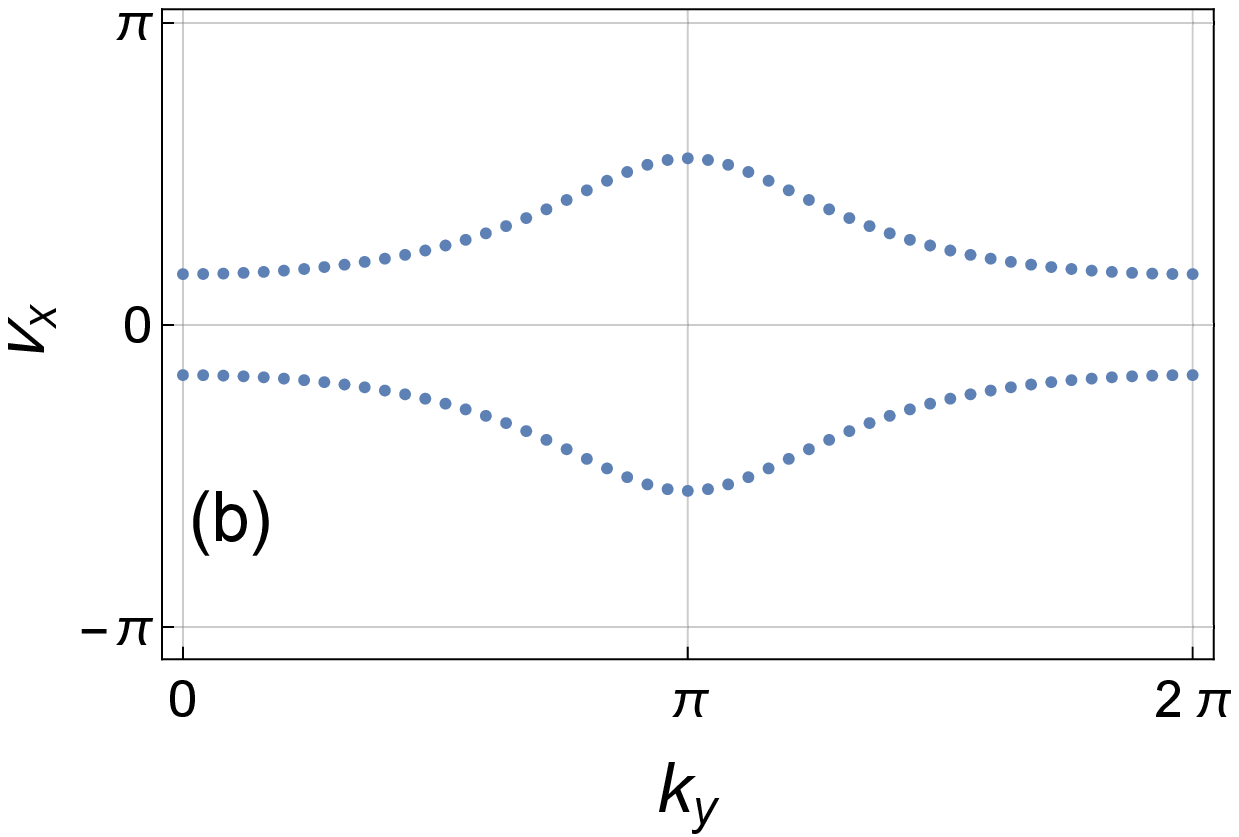}
&
\includegraphics[scale=0.33]{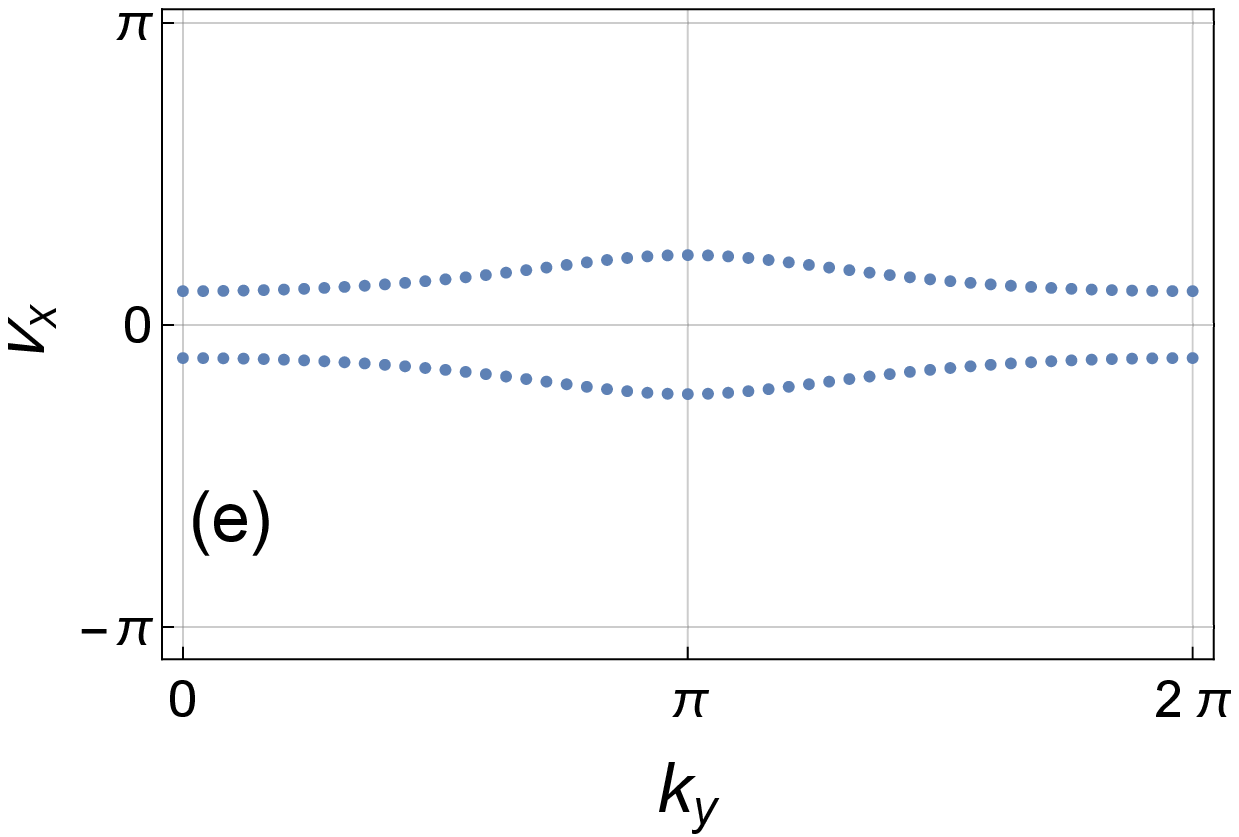}
\\
\includegraphics[scale=0.33]{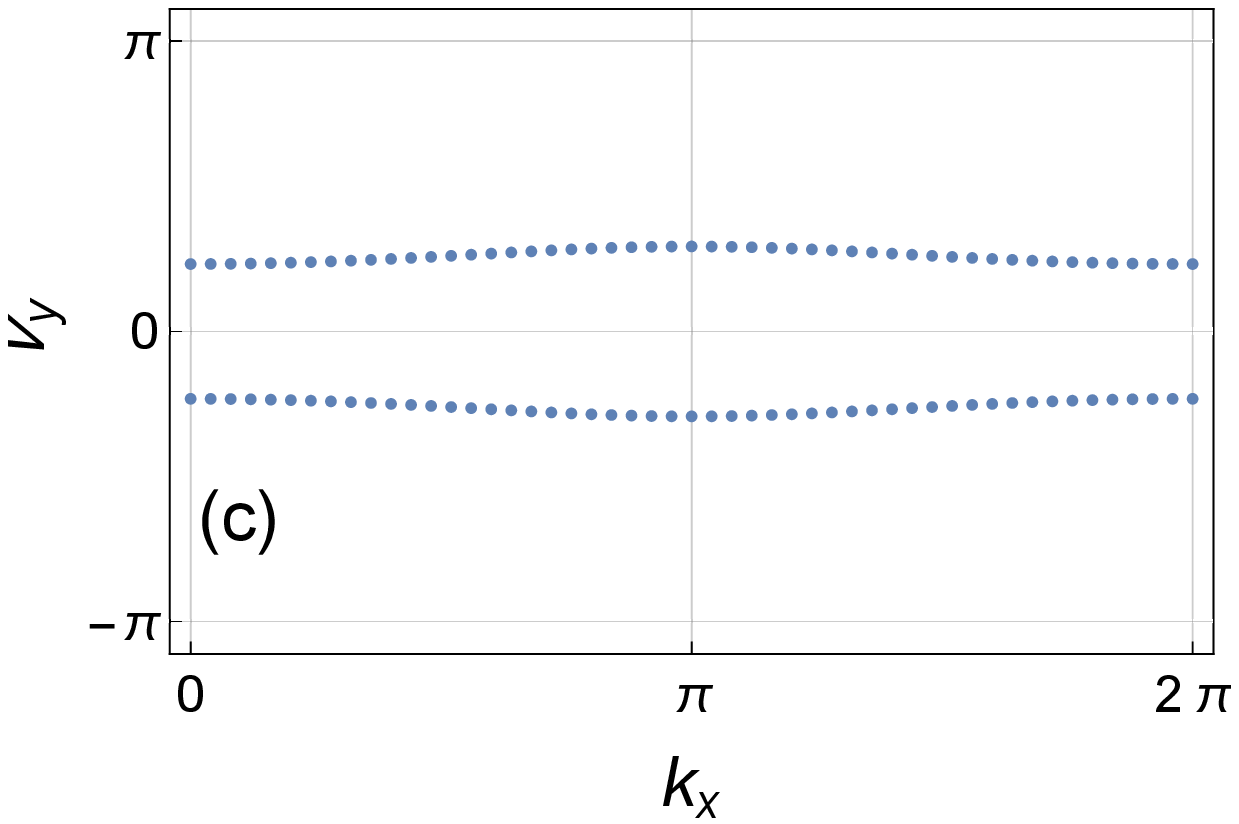}
&
\includegraphics[scale=0.33]{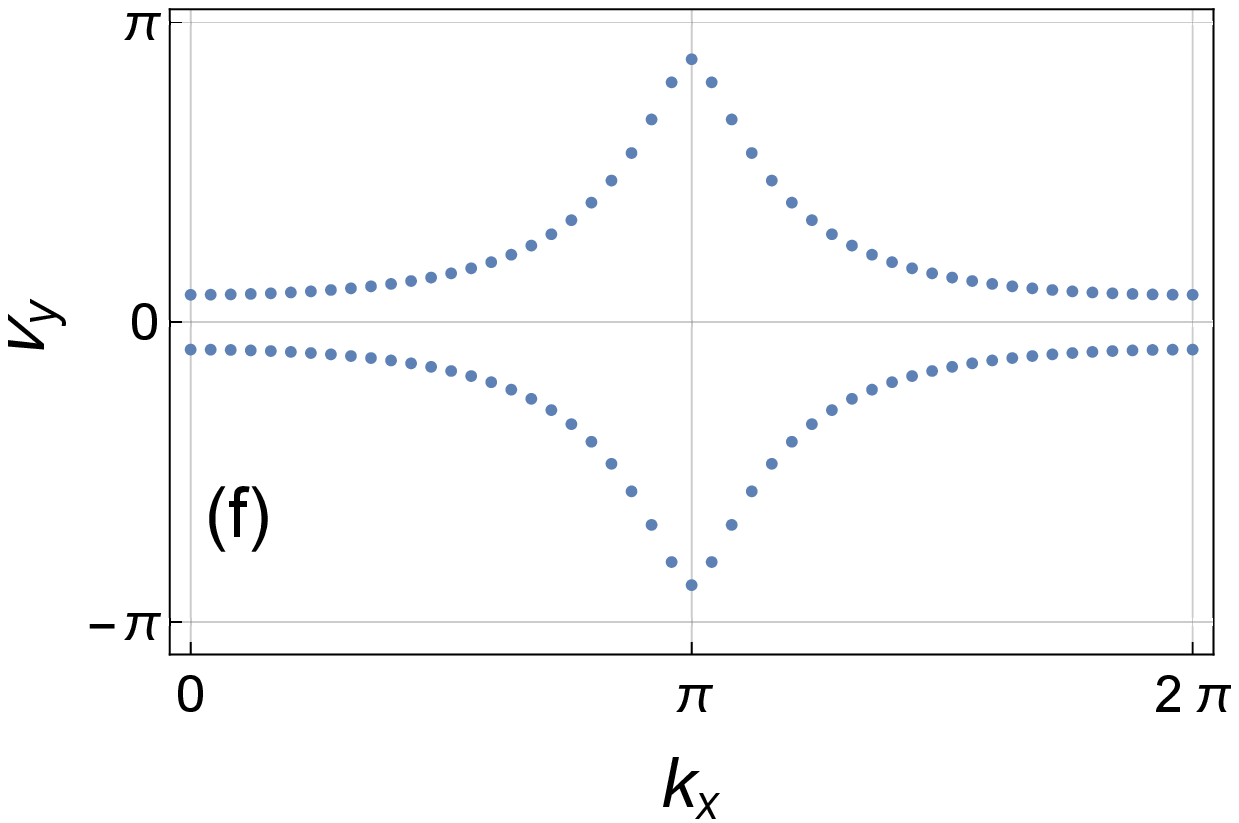}
\end{tabular}
\caption{
(a), (d) Spectra of the Hamiltonian. (b), (e) Wilson loop eigenvalues 
$\nu_x(k_y)$. (c), (f)  Wilson loop eigenvalues $\nu_y(k_x)$.
Left (a), (b),  (c) are models with $\gamma_x=0.1$ and $\gamma_y=0.5$  in the quadrupole $(\pi,\pi)$ phase.
Right (d), (e),  (f) are models with  $\gamma_x=1.1$ and $\gamma_y=0.5$ in the trivial $(0,\pi)$ phase.
Other parameters used are $\lambda_x=\lambda_y=1$.
}
\label{f:wlev}%-----------------------------------------------
\end{center}
\end{figure}

The model Hamiltonian \cite{Benalcazar:2017aa} is defined by
\begin{alignat}1
H=&\sum_j
\big(
\gamma_x c_{3j}^\dagger c_{1j}+\lambda_xc_{1j}^\dagger c_{3j+\hat x}
+\gamma_xc_{2j}^\dagger c_{4j}+\lambda_xc_{4j}^\dagger c_{2j+\hat x}
\nonumber\\
&
-\gamma_yc_{2j}^\dagger c_{3j}-\lambda_yc_{3j}^\dagger c_{2j+\hat y}
+\gamma_yc_{4j}^\dagger c_{1j}+\lambda_yc_{1j}^\dagger c_{4j+\hat y}\big)
\nonumber\\
&+\mbox{H.c.} +\delta\sum_{j,a}(-1)^ac_{aj}^\dagger c_{aj},
\label{Ham}%-------
\end{alignat}
where 
$\gamma_j$ and $\lambda_j$ ($j=x,y$) are hopping parameters toward the $j$ direction,  
$c_{aj}^\dagger$ and $c_{aj}$ are, respectively,  the creation and annihilation operators of electrons at the site $a$ 
in the unit cell labeled by $j=(j_x,j_y)$, as illustrated in Fig. \ref{f:ssh}. 
$\hat j$ ($j=x,y$) stand for the unit vector toward the $j$ direction,
$\hat x=(1,0)$ and $\hat y=(0,1)$.
The minus sign on the bonds connecting 2 and 3 sites show the $\pi$ flux perpendicular to the 2D plane, 
and the last term is a staggered potential  as a symmetry breaking term.
In the momentum representation, we have
\begin{alignat}1
H=\sum_k c_k^\dagger h(k) c_k,
\label{HamMom}%---
\end{alignat}
where 
\begin{alignat}1
h(k) =&(\gamma_x+\lambda_x \cos k_x)\gamma_4+\lambda_x\sin k_x\gamma_3
\nonumber\\
&+(\gamma_y+\lambda_y \cos k_y)\gamma_2+\lambda_y\sin k_y\gamma_1+\delta\gamma_5
\nonumber\\
\equiv&g_\mu(k)\gamma_\mu 
\label{HamEle}%-------
\end{alignat}
with the following $\gamma$-matrices:
\begin{alignat}1
&\gamma_j=-\sigma_j\otimes \sigma_2
=\left(\begin{array}{cc}&i\sigma_j\\-i\sigma_j&\end{array}\right), 
\quad(j=1,2,3)
\nonumber\\
&\gamma_4=\1\otimes \sigma_1
=\left(\begin{array}{cc}&\1\\ \1&\end{array}\right), 
\nonumber\\
&\gamma_5=\1\otimes \sigma_3
=\left(\begin{array}{cc}\1& \\ &-\1\end{array}\right).
\label{GamMat}%-------
\end{alignat}
Here, $\sigma_j$ ($j=1,2,3$) stand for the standard Pauli matrices and $\1$ is the unit matrix.
The notations are similar to those in Ref. \cite{Benalcazar:2017aa}.
Figures \ref{f:wlev}(a) and \ref{f:wlev}(d) are the spectra of the Hamiltonian Eq. (\ref{HamEle}), which shows two doubly degenerate bands.  

As stressed in Ref. \cite{Benalcazar:2017aa}, reflection symmetries with 
respect to the $x$ and $y$ directions play a crucial role in the quadrupole phase of the model:
\begin{alignat}1
&M_x h(k_x,k_y)M_x^{-1}=h(-k_x,k_y),
\nonumber\\
&M_y h(k_x,k_y)M_y^{-1}=h(k_x,-k_y),
\label{RefSym}%-------
\end{alignat}
where $M_x$ and $M_y$ are, respectively,  defined by
\begin{alignat}1
M_x=i\gamma_3\gamma_5=\sigma_3\otimes\sigma_1,\quad M_y=-i\gamma_1\gamma_5=\sigma_1\otimes\sigma_1.
\end{alignat}
Here, we have defined $M_x$ and $M_y$ obeying $M_x^2=M_y^2=1$.
The reflection symmetries ensure the gapped eigenvalues of the U(2) Berry phase (the Wilson loop in the Brillouin zone) 
of the half-filled ground state, as will be  discussed in Sec. \ref{s:Berry}.
This model also has time reversal, particle-hole, and chiral symmetries denoted by
\begin{alignat}1
&Th(k)T^{-1}=h(-k),
\nonumber\\
&Ch(k)C^{-1}=-h(-k),
\nonumber\\
&\gamma_5 h(k)\gamma_5^{-1}=-h(k),
\label{AddSym}%-------
\end{alignat}
where $T=K$ (complex conjugation), $C=K\gamma_5$, and $\gamma_5$ is defined in Eq. (\ref{GamMat}).
When $\gamma_x=\gamma_y$ and $\lambda_x=\lambda_y$, the model has $C_4$ symmetry
\begin{alignat}1
\hat r_4 h(k)\hat r_4^{-1}=h(R_4k),
\end{alignat}
where 
\begin{alignat}1
\hat r_4=
\left(
\begin{array}{cc}
0&1\\-i\sigma_2&0
\end{array}
\right),\quad
R_4\left(\begin{array}{c}k_x\\ k_y\end{array}\right)=\left(\begin{array}{c}k_y\\- k_x\end{array}\right).
\end{alignat}
The spectra in Fig. \ref{f:wlev} are the case with $\gamma_x\ne\gamma_y$, so they  look manifestly 
broken $C_4$-symmetric.

\subsection{Non-Abelian Berry phase}
\label{s:Berry}%---

The BBH model is a 2D version of the 1D SSH model. Therefore, it is natural to characterize topological phases
of the BBH model by means of the Berry phase associated with the 1D SSH model 
\cite{Ryu:2002fk,Hatsugai:2009,Benalcazar61,Benalcazar:2017aa}.
Let 
\begin{alignat}1
\Psi_\pm(k)=\left(\psi_{\pm1}(k),\psi_{\pm2}(k)\right)
\label{GroStaMul}%---
\end{alignat}
be the doubly-degenerate positive and negative energy states of $h(k)$, 
\begin{alignat}1
h(k)\psi_{\pm n}(k)=\pm E(k)\psi_{\pm n}(k),\quad (n=1,2)
\end{alignat}
and let $U_\mu(k)=\Psi_-^\dagger (k)\Psi_-(k+\hat \mu)$ be the 
non-Abelian link variables for the occupied states \cite{FHS05}. 
Here, we assume that $k_\mu$ is discretized  by $N$ mesh points, and
$\hat\mu$ is the unit lattice vector toward the $k_\mu$ direction with length $2\pi/N$.
Then, the $2\times2$ matrix-valued Wilson loop in the Brillouin zone is defined by
\begin{alignat}1
W_\mu(k)=\prod_{n=1}^NU_\mu(k+n\hat \mu).
\label{ConWilLoo}%-------
\end{alignat}
The eigenvalue of the Wilson loop can be parameterized as $e^{i\nu_{\mu}(k)}$, where $\nu_{\mu}(k)$ is associated 
with the polarization.
%\begin{alignat}1
%W_\mu(k)u_{\mu,n}(k)=e^{i\nu_{\mu,n} (k)}u_{\mu,n}(k)
%\label{WilLooEig}%-------
%\end{alignat}
This model shows gapped spectrum $\nu_\mu(k)$, as shown in Fig. \ref{f:wlev}.
They are symmetric with respect to zero,  which implies that two particles 
in a unit cell are polarized in opposite directions, but the total dipole polarization vanishes.

Benalcazar {\it et. al.} \cite{Benalcazar:2017aa}
have introduced the nested Wilson loop, which clarifies the polarization in a projected space of one of the
eigenstate of the Wilson loop. 
In this paper, we will take a different root to reveal the topological property of the BBH model, using the
entanglement topological numbers \cite{Fukui:2014qv,Fukui:2015fk}.

\section{Entanglement polarization}
\label{s:EntPol}%---

Let us divide the system into two subsystems denoted by $A$ and $B$.
If the partition into $A$ and $B$  is spatial and there are boundaries between them, the 
entanglement spectrum (eS) of the reduced density matrix shows fictitious edge states along those boundaries.
%between $A$ and $B$.
This informs us of  the topological properties of the original system \cite{Ryu:2006fk,PhysRevLett.101.010504}.
The eS has been used indeed for the studied of 
noninteracting symmetry-protected topological insulators
\cite{Prodan:2010fj,Turner:2010ys,Hughes:2011rm,Alexandradinata:2011gf,Cirac:2011ul,Huang:2012db,Fang:2013dq}.

Instead of such a spatial partition breaking translational invariance, we can consider another type of partition with respect to some 
internal degrees of freedom, which preserves translational invariance even for subsystems \cite{Hsieh:2014qy,Fukui:2014qv}. 
To be concrete, 
consider a system with a unit cell containing $N$ species such as orbitals, spins, etc., labeled by $i=1,2,\cdots, N$.
Introduce  a partition concerning such species,  $A=\{1,\cdots,n_A\}$ and $B=\{n_A+1,\cdots,n\}$.
Then, the eS remains gapped or becomes gapless. 
This has nothing to do with  the fictitious edge states mentioned above, since the partition does not have any spatial boundaries.
%This is determined, according to 
It depends on whether the partition into $A$ and $B$ is disentangled or entangled that determines such a  spectral property. 
This may be understood as follows:
Let $|G\rangle$ be the many-body ground state. Then, it can be Schmidt-decomposed into
\begin{alignat}1
|G\rangle=\sum_{i,j}D_{ij}|\psi_i\rangle_A\otimes|\phi_j\rangle_B,
\end{alignat} 
where $|\psi_i\rangle_A$ and $|\phi_j\rangle_B$ are orthonormal basis states associated with $A$ and $B$.
The singular-value decomposition of $D=U\Lambda V^\dagger$ , where 
$\Lambda=\mbox{diag}(\lambda_1,\lambda_2,\cdots,\lambda_m,0,\cdots,0)$, leads to
\begin{alignat}1
|G\rangle=\sum_{\ell}\lambda_{\ell}|\Psi_\ell\rangle_A\otimes|\Phi_\ell\rangle_B,
\label{EntGraSta}%---
\end{alignat} 
where $|\Psi_\ell\rangle_A=\sum_i|\psi_i\rangle_AU_{i\ell}$ and $|\Phi_\ell\rangle_B=\sum_j|\phi_j\rangle_BV^*_{j\ell}$, and 
we assume $\lambda_\ell$ in descending order, $(1\ge) \lambda_1\ge\lambda_2\ge\cdots\lambda_m(\ge0)$.
It is known that $\lambda_\ell$ is given by the eS and that if the eS is gapped, the largest $\lambda_1$ is unique.
Then, in this case, let us consider  an  adiabatic deformation of the ground state $|G\rangle$ into
\begin{alignat}1
|G\rangle\sim |\Psi_1\rangle_A\otimes |\Phi_1\rangle_B.
\label{TenPro}%---
\end{alignat}
It induces 
%by making 
the gap of the eS to be larger. In this process, 
the topological properties of 
$|\Psi_1\rangle_A$ and $|\Phi_1\rangle_B$ are unchanged due to the gap.
Thus, it can be useful to characterize $|G\rangle$ in terms of the topological properties of 
$|\Psi_1\rangle_A$ and $|\Phi_1\rangle_B$.
These states are the ground states of the eHs, which will be defined below in Sec. \ref{s:def_eP}.

Let $c_{G}$ be a topological invariant such as the Chern number or dipole polarization of the ground state $|G\rangle$.
Then, Eq. (\ref{TenPro}) implies the partition of the topological number $c_G$ into
$c_G=c_A+c_B$, where $c_A$ and $c_B$
are the topological numbers associated with $A$ and $B$, which may be referred to 
as entanglement topological numbers \cite{Fukui:2015fk}.

As we will show below, when the above scenario 
is applied to the polarization, it enables us to distinguish whether the system has the quadrupole polarization or not.
Namely, even though the quadrupole phase has vanishing total dipole polarization, 
if part of the dipole polarizations among the quadrupole polarization is traced out, the other dipole polarization is
revealed. 

Below, we introduce the eH for an appropriate partition of the quadrupole phase, and using the eigenstates of the eH,
we argue the eP for the BBH model.

%The entanglement spectrum (eS) of the reduced density matrix informs us of the edge states along artificial boundaries introduced 
%by a partition of the system \cite{Ryu:2006fk,PhysRevLett.101.010504}.
%The eS has been used for the studied of 
%noninteracting symmetry-protected topological insulators
%\cite{Prodan:2010fj,Turner:2010ys,Hughes:2011rm,Alexandradinata:2011gf,Cirac:2011ul,Huang:2012db,Fang:2013dq}.
%We show below that not only the spectrum but also a wave function of the eH is useful to reveal topological properties
%of the system. To this end, we introduce an extensive partition which keeps the translational invariance for the eH 
%\cite{Hsieh:2014qy,Fukui:2014qv}.
%This enables us to define the topological invariant for the eH of the subsystem.

\subsection{Definition of entanglement polarization}
\label{s:def_eP}

Let $|G\rangle$ be the many-body ground state, and let $\rho=|G\rangle\langle G|$ be the corresponding density matrix.
We divide the four sites labeled by $a,b=1,\cdots,4$ in the unit cell into a pair $i,j$ and its complement $\bar i,\bar j$. 
Then, tracing out one pair, 
we have the eH ${\cal H}^{(ij)}$ and $\bar{\cal H}^{(ij)}$,
\begin{alignat}1
&\tr_{\bar i,\bar j}~\rho\propto e^{-{\cal H}^{(ij)}},
\nonumber\\
&\tr_{i,j}~\rho\propto e^{-\bar{\cal H}^{(ij)}},
\end{alignat}
where we have simply denoted ${\cal H}^{(\bar i\bar j)}=\bar{\cal H}^{(ij)}$.
It has been shown that for non-interacting fermions, the eH becomes also noninteracting,
${\cal H}^{(ij)}=\sum_{(a,b)\in(i,j)}c_a^\dagger h_{ab}^{(ij)}c_b$ and
$\bar{\cal H}^{(ij)}=\sum_{(a,b)\in(\bar i,\bar j)}c_a^\dagger \bar h_{ab}^{(ij)}c_b$.
It has a simple relationship with the correlation function \cite{Peschel:2003uq}.

Let us discuss the eigenvalue problem for the eH with translational symmetry.
Using the negative energy multiplet wave function Eq. (\ref{GroStaMul}),
the projection operator $P_-(k)$ at a given $k$ to the negative energy states is defined by
\begin{alignat}1
P_-(k)\equiv \Psi_-(k)\Psi_-^\dagger(k),
\label{ProOpe}%-------
\end{alignat}
where $\Psi_-(k)$ is the ground state (negative energy) multiplet wave function in Eq. (\ref{GroStaMul}).
Note that this is a $4\times 4$ matrix $P_-(k)=P_{-,ab}(k)$ $(a,b=1,2,3,4)$ associated with four sites in the unit cell.
We define the projection operator restricted to particular two sites $i,j$ by
\begin{alignat}1
P_-^{(ij)}(k)\equiv P^{(ij)}P_-(k)P^{(ij)},
\label{EntProOpe}%---
\end{alignat}
where $P^{(ij)}$ is the projection to the space spanned by $i$ and $j$.
For example, $P^{(13)}=\mbox{diag}(1,0,1,0)$.
Equivalently, it can be expressed as $P^{(ij)}_{-,ab}(k)=P_{-,ab}(k)$ for $(a,b)=(i,j)$, 
which is a reduced $2\times2$ matrix. In this reduced form, 
trivial 0 eigenvalues of the $4\times4$ projection operator are removed.
Likewise, $\bar P_-^{(ij)}(k)\equiv P_-^{(\bar i\bar j)}(k)$ is defined.

Instead of the eigenvalue problem of the eH,
let us consider that of the reduced projection operator of Eq. (\ref{EntProOpe}) introduced above,
\begin{alignat}1
P^{(ij)}_-(k)\psi^{(ij)}_n(k)=\xi^{(ij)}_n(k)\psi^{(ij)}_n(k),\quad (n=1,\cdots,M),
\label{EntEigVal}%-------
\end{alignat}
where in the present case, $M=2$.
It has been shown
that the eigenvector $\psi_n^{(ij)}(k)$ is simultaneously the eigenvector of eH $h^{(ij)}(k)$ in the
momentum representation, $h^{(ij)}(k)\psi^{(ij)}_n(k)=\varepsilon^{(ij)}_n(k)\psi^{(ij)}_n(k)$ \cite{Peschel:2003uq}.
The relationship between the eigenvalues is $\xi^{(ij)}(k)=\big[\exp\big(\varepsilon^{(ij)}(k)\big)+1)\big]^{-1}$ 
\cite{Peschel:2003uq}.
Therefore, the eigenvalue $\xi^{(ij)}_n$ will be simply referred to as eS.
It should be noted that the spectrum of $P_-^{(ij)}(k)$ and its complement  $\bar P_-^{(ij)}(k)$ have the following relationship
\begin{alignat}1
\bar\xi_n^{(ij)}(k)=1-\xi_{M+1-n}^{(ij)}(k),
%\quad ( m\ne n ),
\label{EntSpeCom}%---
\end{alignat}
where $\bar\xi_n^{(ij)}(k)$ is the eigenvalue of $\bar P_-^{(ij)}(k)$.
Namely, an occupied (unoccupied) state of $P_-^{(ij)}(k)$ corresponds to an unoccupied (occupied) state of 
$\bar P_-^{(ij)}(k)$.

When the eS is gapped at $\xi=1/2$, 
\begin{alignat}1
\xi_1^{(ij)}(k)<1/2<\xi_2^{(ij)}(k),
\label{GapEntSpe}%---
\end{alignat}
we can define the entanglement Wilson loop in the Brillouin zone (eWL) 
associated with the occupied state labeled by $n=2$.
To this end, we first introduce the entanglement link variable as
${\cal U}_\mu^{(ij)}(k)=\psi_2^{(ij)\dagger}(k)\psi_2^{(ij)}(k+\hat \mu)$, 
and next define the eWL as
\begin{alignat}1
{\cal W}^{(ij)}_\mu(k)=\prod_{n=1}^N {\cal U}_\mu^{(ij)}(k+n\hat\mu).
\label{EntWilLoo}%-------
\end{alignat} 
This leads to the idea of the eP ${\cal P}^{(ij)}_\mu(k)$ as the phase of the eWL,
\begin{alignat}1
{\cal P}^{(ij)}_\mu(k)=\arg \big[{\cal W}^{(ij)}_\mu(k)\big].
\label{EntPol}%-------
\end{alignat}

\subsection{Symmetry property of eP and eS}

Let us next consider the symmetry properties of eP and eS.
The projection operator $P_-(k)$ obeys the same symmetry property of the Hamiltonian $h(k)$ in 
Eqs. (\ref{RefSym}) and (\ref{AddSym}).
Then, the following relationship
\begin{alignat}1
&M_xP^{(13)}M_x^{-1}=P^{(13)},\quad M_yP^{(13)}M_y^{-1}=P^{(24)},
\nonumber\\
&M_xP^{(14)}M_x^{-1}=P^{(23)},\quad M_yP^{(14)}M_y^{-1}=P^{(14)},
\nonumber\\
&M_xP^{(12)}M_x^{-1}=P^{(34)},\quad M_yP^{(12)}M_y^{-1}=P^{(34)},
\label{ProRef}%---
\end{alignat}
leads to 
\begin{alignat}1
&M_x P^{(13)}_-(k_x,k_y)M_x^{-1}=P^{(13)}_-(-k_x,k_y),
\nonumber\\
&M_yP^{(14)}_-(k_x,k_y)M_y^{-1}=P^{(14)}_-(k_x,-k_y).
\label{RefSymEnt}%-------
\end{alignat}
These guarantee the quantization of the eP, 
\begin{alignat}1
&{\cal P}_x^{(13)}(k)\overset{M_x}{=}0\mbox{ or }\pi,
\nonumber\\
&{\cal P}_y^{(14)}(k)\overset{M_y}{=}0\mbox{ or }\pi,\quad \mbox{mod }2\pi.
\end{alignat}
If the gap of the eH is open in the whole Brillouin zone, these values are constant 
independent of  $k$.
Therefore, we propose that the pair of the eP,
\begin{alignat}1
({\cal P}^{(13)}_x,{\cal P}^{(14)}_y)
\label{BulEntPol}%-------
\end{alignat}
characterize the phases of the present model.

The eS also has some symmetry properties. The reflection symmetries Eqs. (\ref{RefSymEnt}) lead to
\begin{alignat}1
&\xi_n^{(13)}(k_x,k_y)\overset{M_x}{=}\xi^{(13)}_n(-k_x,k_y),
\nonumber\\
&\xi_n^{(14)}(k_x,k_y)\overset{M_y}{=}\xi^{(14)}_n(k_x,-k_y).
\label{RefSymCon}%---
\end{alignat}
Equation (\ref{ProRef}) yields another relationship between $P_-^{(ij)}(k)$, 
\begin{alignat}1
M_yP^{(13)}_-(k_x,k_y)M_y^{-1}= P^{(24)}_-(k_x,-k_y),
\nonumber\\
M_xP^{(14)}_-(k_x,k_y)M_x^{-1}= P^{(23)}_-(-k_x,k_y).
\end{alignat}
These lead to
\begin{alignat}1
&\xi_n^{(13)}(k_x,k_y)\overset{M_y}{=}\xi^{(24)}_n(k_x,-k_y),
\nonumber\\
&\xi_n^{(14)}(k_x,k_y)\overset{M_x}{=}\xi^{(23)}_n(-k_x,k_y).
\label{RefCro}%---
\end{alignat}
Note that the right-hand side are the eS of the complement partitions of the left-hand side.
For example, $\xi^{(24)}_n(k)=\bar\xi^{(13)}_n(k)$. 
Thus, 
combining these with Eq. (\ref{EntSpeCom}), we have
\begin{alignat}1
&\xi_n^{(13)}(k_x,k_y)+\xi_{M+1-n}^{(13)}(k_x,-k_y)\overset{M_y}{=}1,
\nonumber\\
&\xi_n^{(14)}(k_x,k_y)+\xi_{M+1-n}^{(14)}(-k_x,k_y)\overset{M_x}{=}1,
\label{RefSpeSym}%---
\end{alignat}
where $M=2$ in the present system.

Chiral symmetry in $P^{(ij)}_-(k)$ is implemented by 
\begin{alignat}1
\gamma_5 P_-^{(ij)}(k)\gamma_5=P^{(ij)}\big[1-P_-^{(ij)}(k)\big]P^{(ij)}.
%=P^{(ij)}(1-P_-^{(ij)}(k))P^{(ij)}
\end{alignat}
In the reduced $2\times2$ representation, it becomes $\gamma_5 P_-^{(ij)}(k)\gamma_5=1-P_-^{(ij)}(k)$
with $\gamma_5=\mbox{diag}(1,-1)$ for $(ij)=(13), (14)$, whereas $\gamma_5=\mbox{diag}(1,1)$ for $(ij)=(12)$.
%On the other hand, chiral symmetry gives the constraint
Thus, provided that the eS is gapped Eq. (\ref{GapEntSpe}), $\gamma_5$ converts the occupied band into the unoccupied band
and vice versa for $(ij)=(13), (14)$.
Considering the fact that total bands (sum of the occupied band and unoccupied band)
give a trivial polarization and that $\gamma_5$ does not change the polarization,
we have
\begin{alignat}1
{\cal P}^{(ij)}_\mu(k)\overset{\gamma_5}{=}0\mbox{ or }\pi, \quad \mbox{ mod }2\pi,
\end{alignat}
and
\begin{alignat}1
\xi_n^{(ij)}(k)+\xi_{M+1-n}^{(ij)}(k)\overset{\gamma_5}{=}1,
\end{alignat} 
for $(ij)=(13), (14)$, and their complements. Together with Eq. (\ref{EntSpeCom}), we have
\begin{alignat}1
\xi_n^{(ij)}(k)\overset{\gamma_5}{=}\bar\xi_n^{(ij)}(k),
\end{alignat}
for $(ij)=(13), (14)$, and their complements. Further together with Eq. (\ref{RefCro}), we conclude
\begin{alignat}1
&\xi_n^{(13)}(k_x,k_y)\overset{M_y,\gamma_5}{=}\xi_n^{(13)}(k_x,-k_y),
\nonumber\\
&\xi_n^{(14)}(k_x,k_y)\overset{M_x,\gamma_5}{=}\xi_n^{(14)}(-k_x,k_y).
\label{ChiAddSym}%---
\end{alignat}
In passing, we mention that $P^{(12)}_-(k)\overset{\gamma_5}{=}1/2$.

Time reversal symmetry yields
\begin{alignat}1
{\cal P}^{(ij)}_\mu(k)\overset{T}{=}{\cal P}^{(ij)}_\mu(-k),\quad \mbox{ mod }2\pi,
\end{alignat}
and 
\begin{alignat}1
\xi^{(ij)}_n(k)\overset{T}{=}\xi^{(ij)}_n(-k).
\label{TimAddSym}%---
\end{alignat}
The constraint of particle-hole symmetry is just that of the combination of time reversal and chiral symmetries
$C=T\gamma_5$.

Finally let us  briefly discuss $C_4$ symmetry. Note the following transformation property of $P^{(ij)}$,
\begin{alignat}1
&\hat r_4P^{(13)}\hat r_4^{-1}=P^{(14)},\quad \hat r_4P^{(14)}\hat r_4^{-1}=P^{(24)},
\nonumber\\
&\hat r_4P^{(24)}\hat r_4^{-1}=P^{(23)},\quad \hat r_4P^{(23)}\hat r_4^{-1}=P^{(13)},
\nonumber\\
&\hat r_4P^{(12)}\hat r_4^{-1}=P^{(34)},\quad \hat r_4P^{(34)}\hat r_4^{-1}=P^{(12)}.
\end{alignat}
Together with
\begin{alignat}1
\hat r_4P_-(k)\hat r_4^{-1} =P_-(R_4k),
\end{alignat}
we reach
\begin{alignat}1
\hat r_4 P_-^{(13)}(k) \hat r_4^{-1}=P_-^{(14)}(R_4k).
\end{alignat}
This relation leads to
\begin{alignat}1
&{\cal P}_x^{(14)}(k_y)\overset{C_4}{=}{\cal P}_y^{(13)}(k_x=-k_y),
\nonumber\\
&\xi_n^{(14)}(k_x,k_y)\overset{C_4}{=}\xi_n^{(13)}(-k_y,k_x).
\end{alignat}

\subsection{Exact eP for the BBH model}
\label{s:Exact}%---

The BBH model Eq. (\ref{Ham}) allows an exact result for the eP.
The projection operator Eq. (\ref{ProOpe}) is given by
\begin{alignat}1
P_-(k)=\frac{1}{2}\Big(1-\frac{h(k)}{E(k)}\Big),
\label{ProOpeMin}%---
\end{alignat}
where $E(k)=\sqrt{g_\mu^2(k)}$. Thus, except for the case $|\gamma_\mu|=|\lambda_\mu|$ for both $\mu=x,y$, 
the spectrum of the model is gapped, and Eq. (\ref{ProOpeMin}) is well-defined.
We then have 
\begin{alignat}1
&P^{(13)}_-(k)=\frac{1}{2}
\left(\begin{array}{cc}1-\tilde g_5&-\tilde g_4-i\tilde g_3\\ -\tilde g_4+i\tilde g_3 &1+\tilde g_5\end{array}\right),
\nonumber\\
&P^{(14)}_-(k)=\frac{1}{2}
\left(\begin{array}{cc}1-\tilde g_5&-\tilde g_2-i\tilde g_1\\ -\tilde g_2+i\tilde g_1 &1+\tilde g_5\end{array}\right),
%\nonumber\\
%&P^{(12)}_-(k)=\frac{1}{2}(1-\tilde g_5)\1,
\label{EffSSH}%-------
\end{alignat}
where $\tilde g_\mu(k)=g_\mu(k)/E(k)$.
Since $P^{(13)}_-(k)$ and $P^{(14)}_-(k)$ in Eq. (\ref{EffSSH}) are  basically 
independent 1D SSH Hamiltonians along $x$ and $y$ directions, respectively, 
it turns out that when $g_5=0$, the winding number of $g_4+ig_3$ and $g_2+ig_1$ determine the eP in Eq. (\ref{BulEntPol}).
Namely, 
${\cal P}_x^{(13)}=\pi$ for $|\gamma_x/\lambda_x|< 1$ and ${\cal P}_x^{(13)}=0$ otherwise,
and ${\cal P}_y^{(14)}=\pi$ for $|\gamma_y/\lambda_y|<1$ and ${\cal P}_x^{(13)}=0$ otherwise.
The quadrupole phase is characterized by $(\pi,\pi)$ as nontrivial SSH phase both for the $x$ and $y$ directions.
This feature can be clearly seen by the eP of the edge states, as will be discussed in the next section.

The bulk eP is also expressed by the eigenvalues of reflection operators $M_x$ and $M_y$, because they commute with 
$P^{(13)}_-(k)$  and $P^{(14)}_-(k)$, respectively, at the reflection-invariant lines $(k_x^*,k_y)$ and $(k_x,k_y^*)$, where
$k_x^*,k_y^*\equiv 0, \pi$
\cite{Benalcazar:2017aa,Hughes:2011rm}. 
Let $p^{(13)}_{k_x^*}$ and $p^{(14)}_{k_y^*}$ be the eigenvalues of $M_x$ and $M_y$, respectively, 
on the invariant lines specified by $k_x^*$ and $k_y^*$
of the occupied states $\psi^{(13)}_2(k_x^*,k_y)$ and $\psi^{(14)}_2(k_x,k_y^*)$.
Note that in the reduced $(13)$-space where $P_-^{(13)}(k)$ is expressed by a $2\times2$ matrix, $M_x$ acts it as $M_x^{(13)}=\sigma_1$. 
Thus, $p^{(13)}_0p^{(13)}_\pi={\rm sgn }\big[(\gamma_x+\lambda_x)(\gamma_x-\lambda_x)\big]$.
Likewise, in the reduced (14)-space, $M_y$ acts it also as $M_y^{(14)}=\sigma_1$. 
Thus, we finally  reach 
\begin{alignat}1
e^{i{\cal P}_x^{(13)}}=p^{(13)}_0p^{(13)}_\pi,\quad
e^{i{\cal P}_y^{(14)}}=p^{(14)}_0p^{(14)}_\pi.
\label{RefEig}%---
\end{alignat}
We will show in Sec. \ref{s:GenBBH}  that this relation is valid for more generic model.

\section{Entanglement Edge state polarization}
\label{s:eESP}%---

\begin{figure}[htb]
\begin{center}
\includegraphics[scale=0.4]{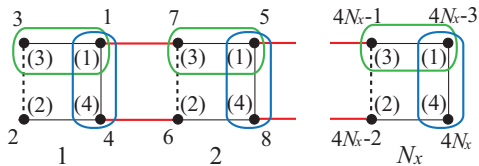}
\caption{
Numbering of the sites for 
the BBH model with open boundaries in the $x$ direction.
The sites encircled by green and blue curves are  associated with 
the partition (13) and (14), respectively.
}
\label{f:open_x}%-----------------------------------------------
\end{center}
\end{figure}

The quadrupole phase is characterized by  the corner edge states. In this sense, the topological quadrupole phase
is nowadays referred to as a second order topological insulator.
In this section, we investigate the topological phase of the present model from the point of view of the edge states.

\begin{figure}[htb]
\begin{center}
\begin{tabular}{cc}
\includegraphics[scale=0.33]{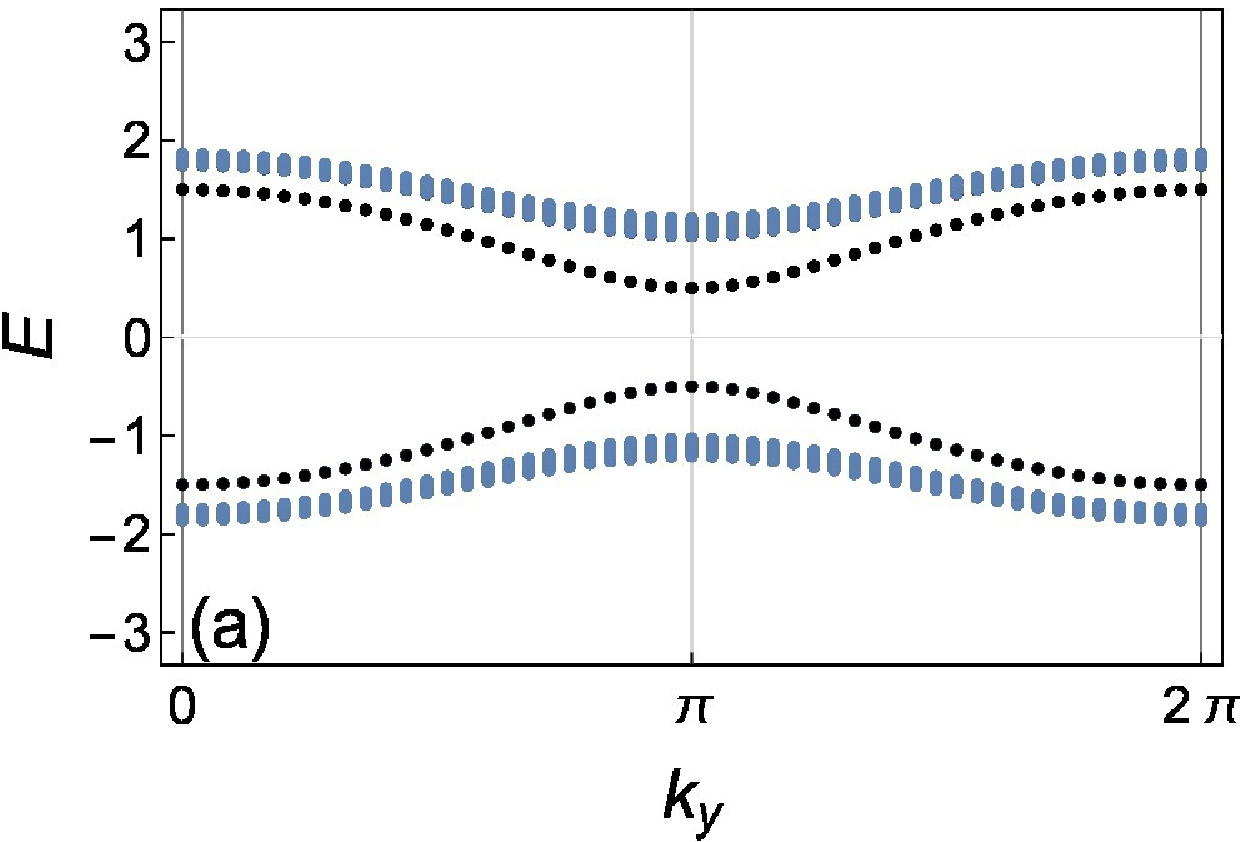}
&
\includegraphics[scale=0.33]{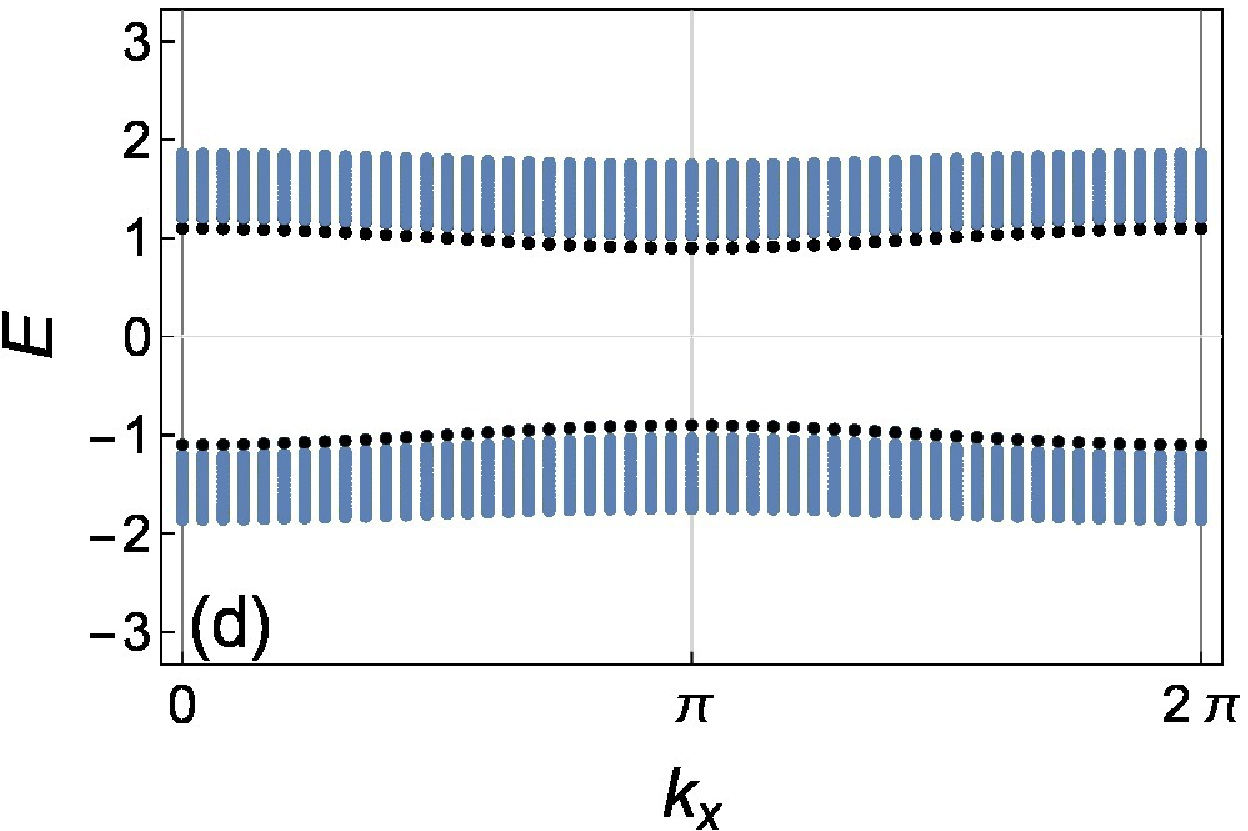}
\\
\includegraphics[scale=0.33]{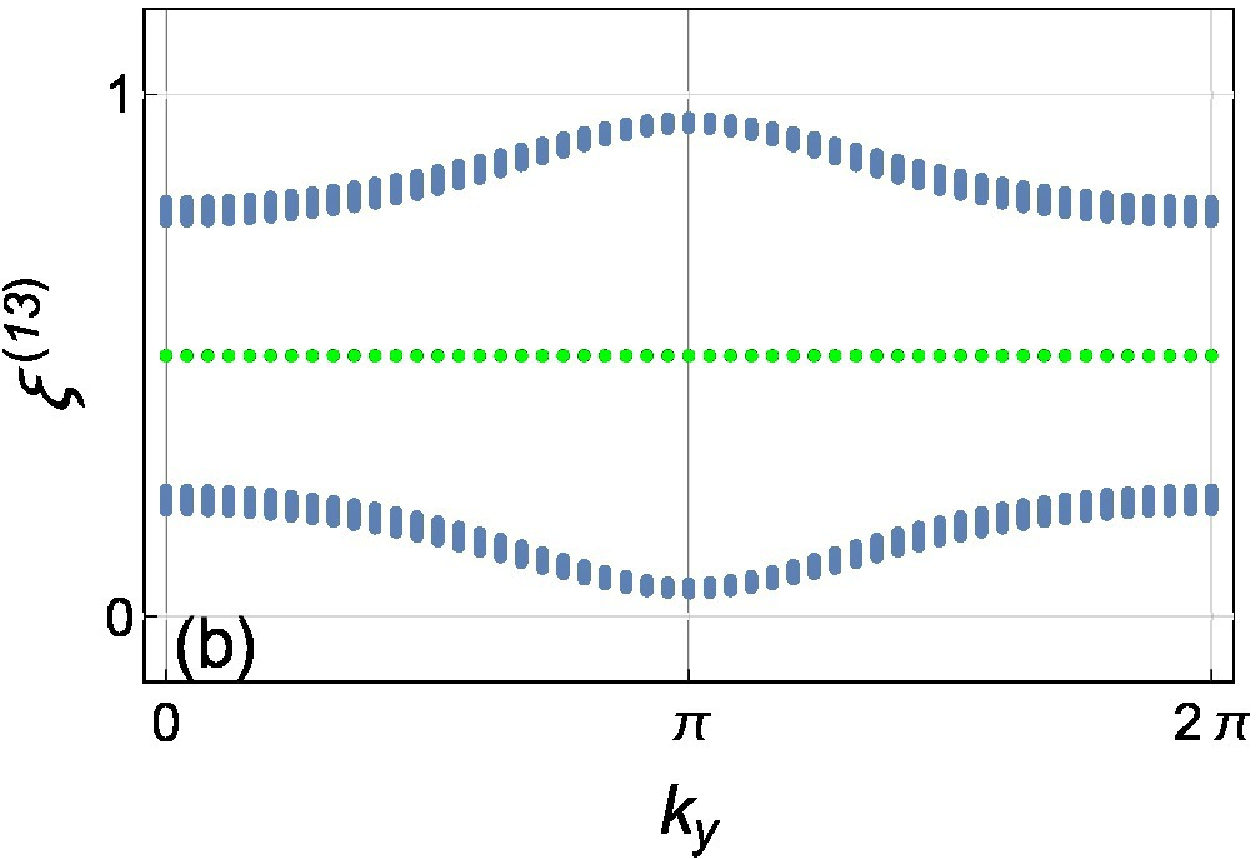}
&
\includegraphics[scale=0.33]{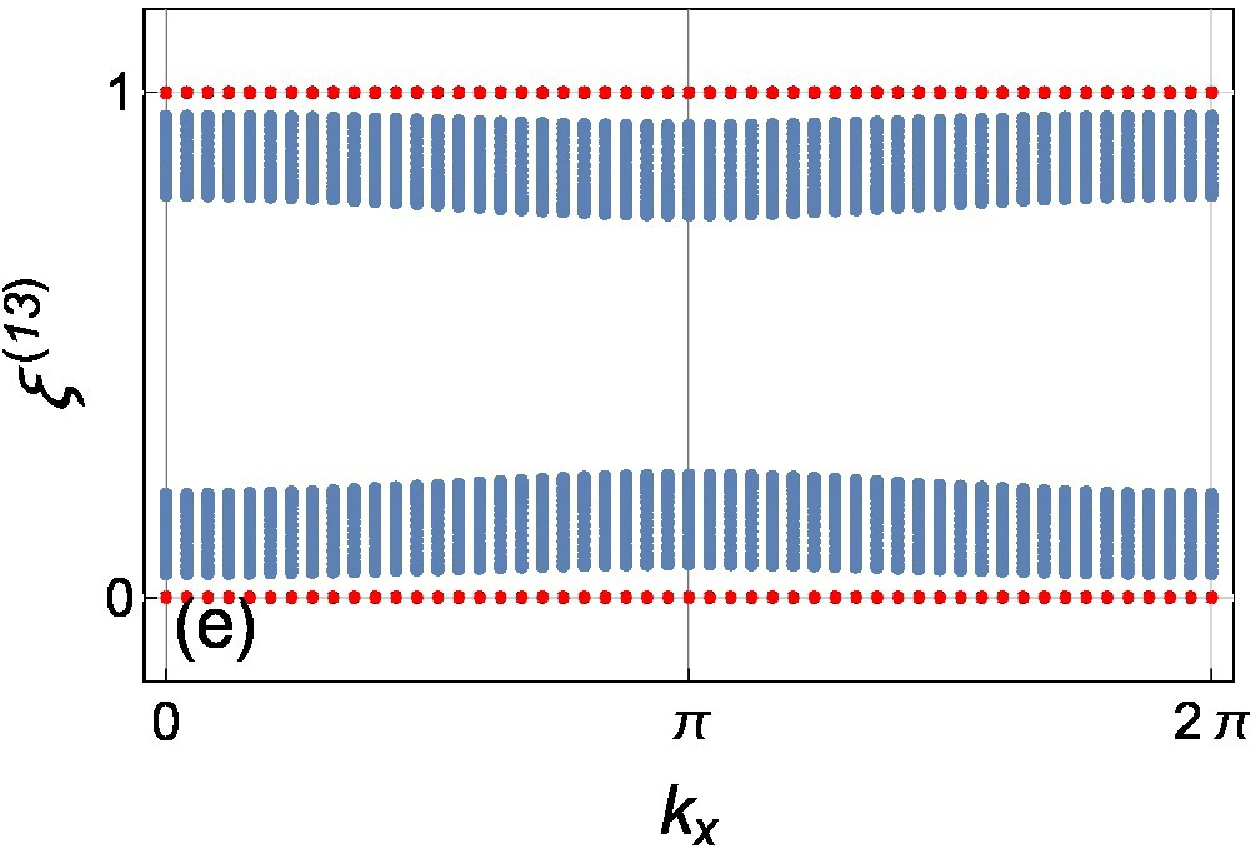}
\\
\includegraphics[scale=0.33]{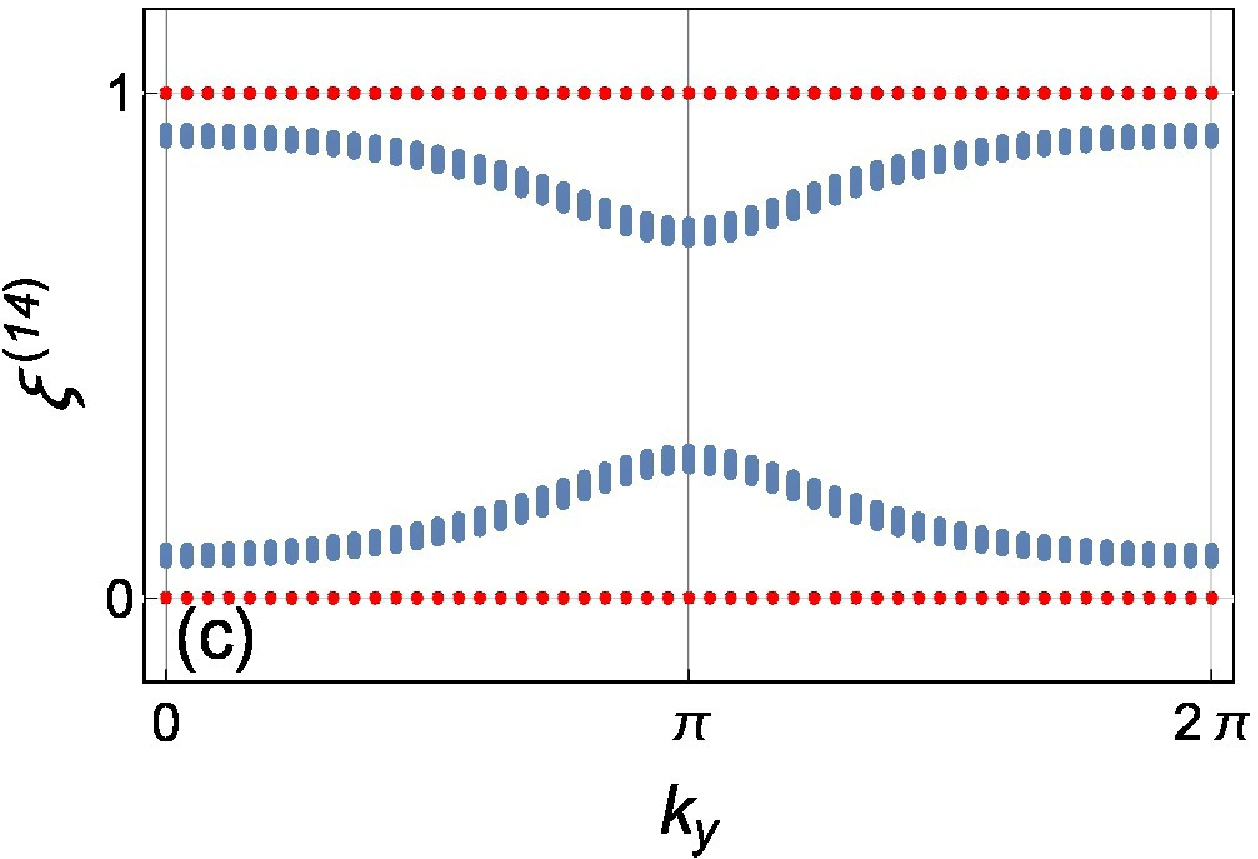}
&
\includegraphics[scale=0.33]{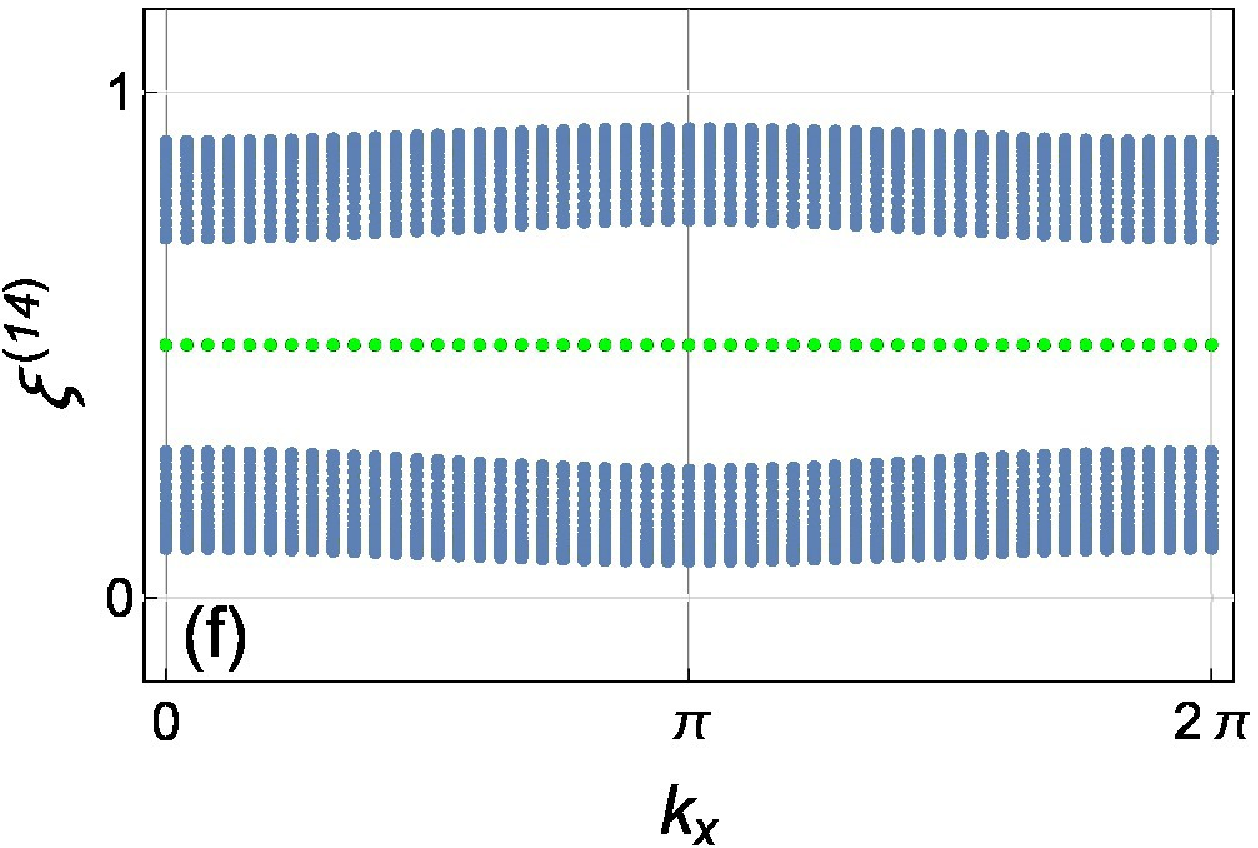}
\end{tabular}
\caption{
Various spectra of a finite system with boundaries in the quadrupole $(\pi,\pi)$ phase ($\gamma_x=0.1$ and $\gamma_y=0.5$). 
(a) Spectrum of the system with open boundaries in the $x$ direction and periodic along the $y$ direction.
(b) eS, $\xi^{(13)}$, and
(c) eS, $\xi^{(14)}$, corresponding to (a).
(d)  Spectrum of the system with open boundaries in the $y$ direction and periodic along the $x$ direction.
%(a), (d) Spectra of the systems with open boundaries. 
(e) eS, $\xi^{(13)}$, and
(f) eS, $\xi^{(14)}$, corresponding to (d).
In (b) and (f), the green lines denote the doubly degenerate zero energy states, and in (c) and (e), the red line shows the 
edge states (fully occupied and fully unoccupied states) with eESP, ${\cal P}_{x\mbox{\scriptsize-edge},y}^{(14)}=\pi$ and
${\cal P}_{y\mbox{\scriptsize-edge},x}^{(13)}=\pi$. 
}
\label{f:edge_qp}%-----------------------------------------------
\end{center}
\end{figure}

\subsection{Gapped edge states}
\label{s:GappedEdge}%---

\begin{figure}[htb]
\begin{center}
\begin{tabular}{cc}
\includegraphics[scale=0.33]{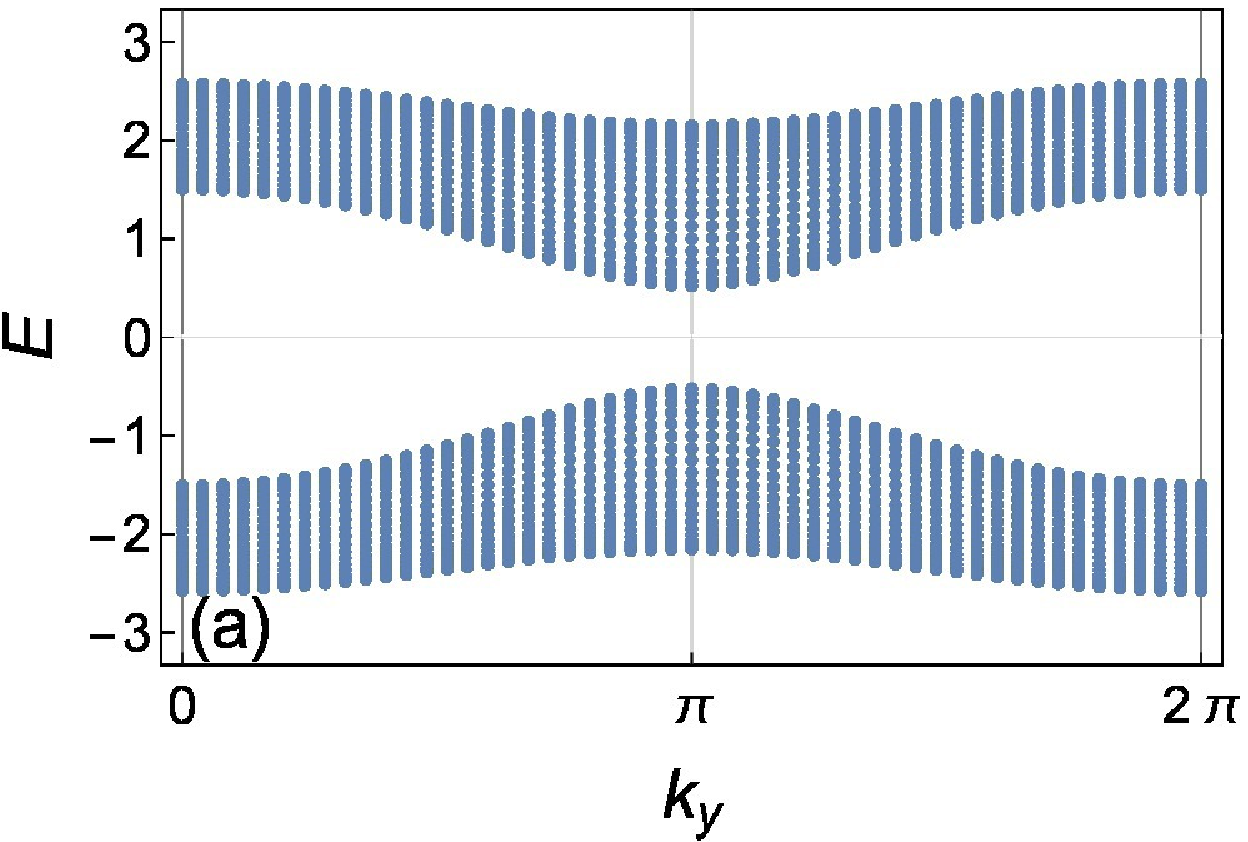}
&
\includegraphics[scale=0.33]{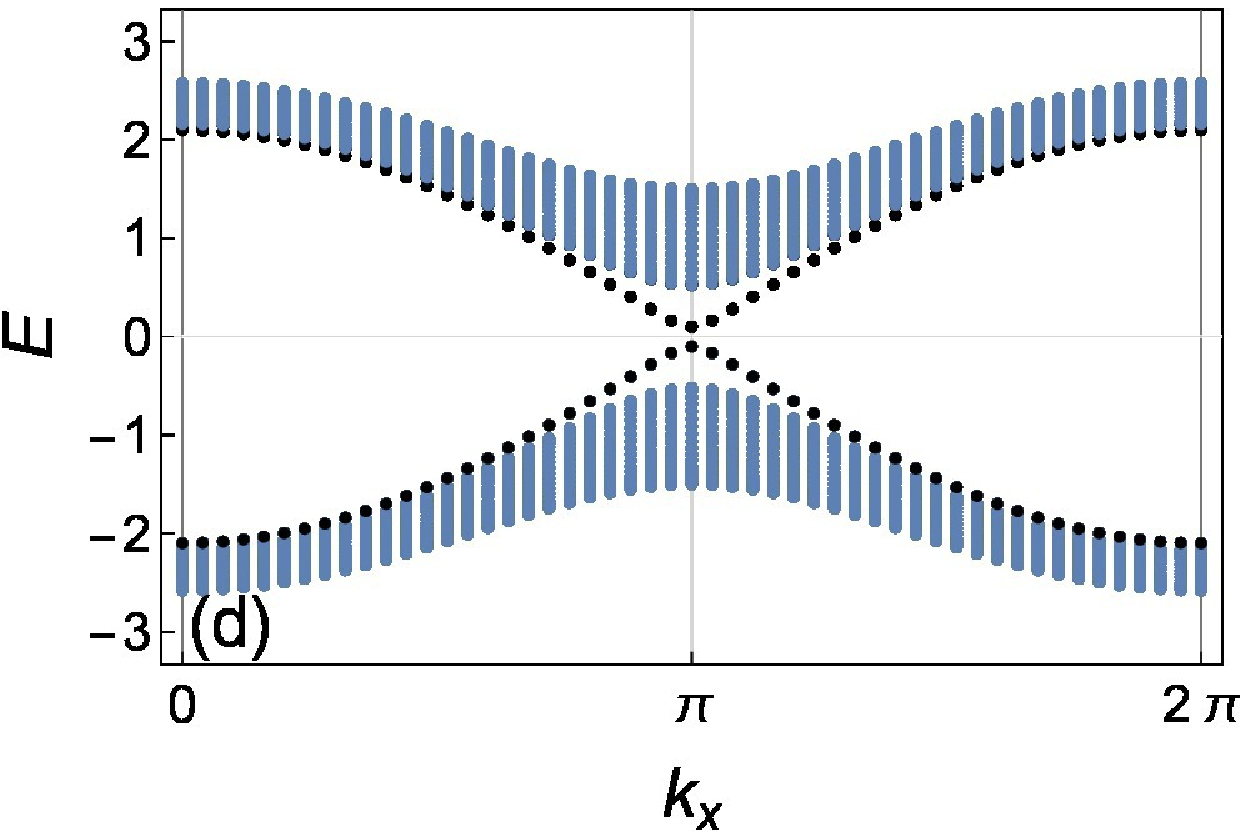}
\\
\includegraphics[scale=0.33]{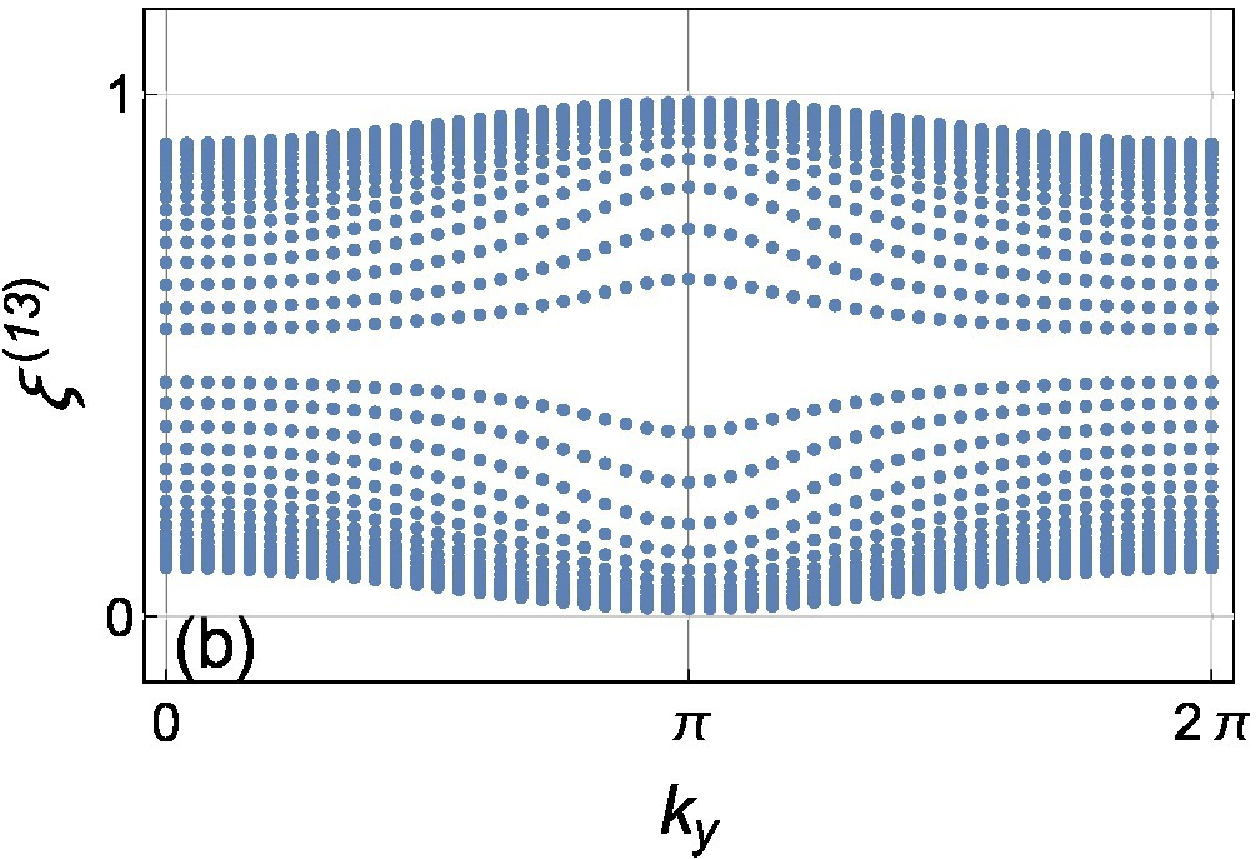}
&
\includegraphics[scale=0.33]{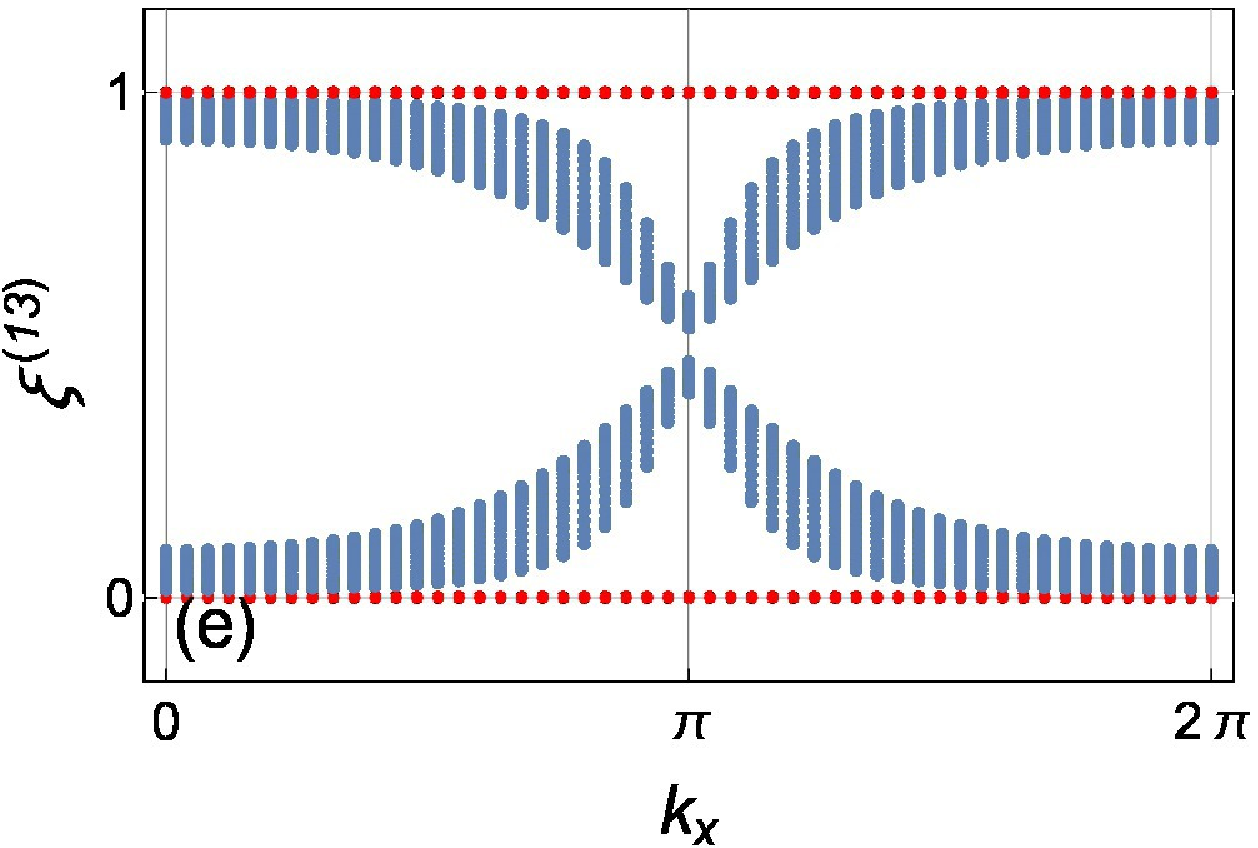}
\\
\includegraphics[scale=0.33]{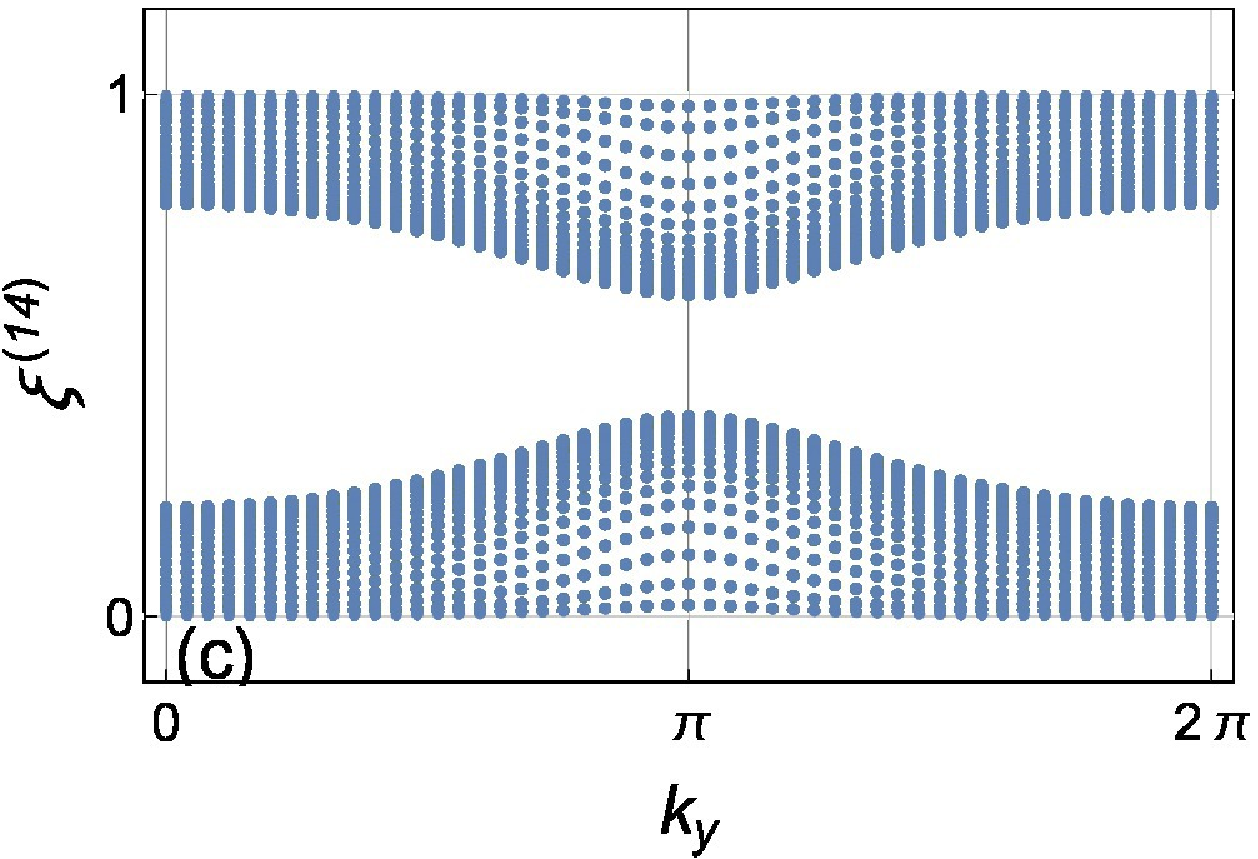}
&
\includegraphics[scale=0.33]{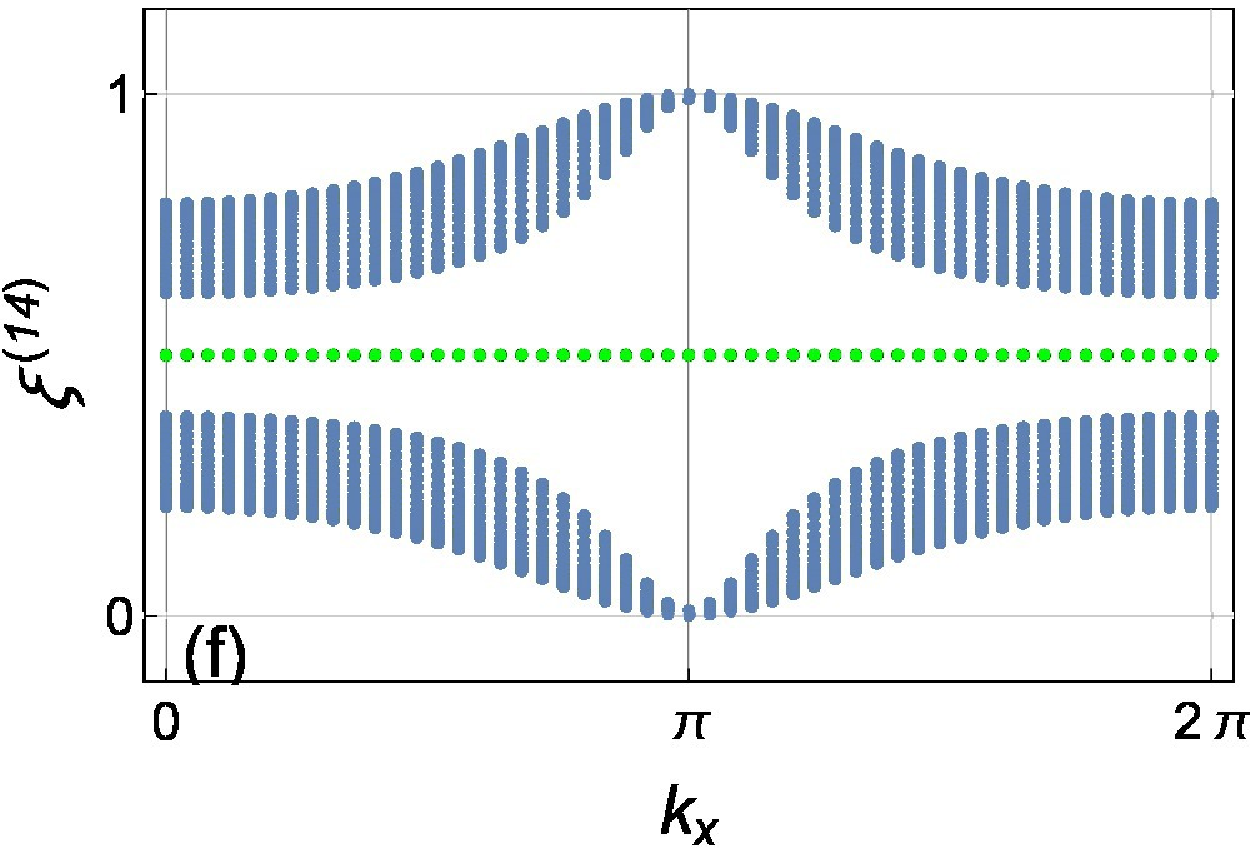}
\end{tabular}
\caption{
The same figures as in Fig. \ref{f:edge_qp}, but
in the trivial $(0,\pi)$ phase ($\gamma_x=1.1$ and $\gamma_y=0.5$). 
In (e), the eES denoted by red lines at $\xi=0$ and $\xi=1$ have eESP, ${\cal P}_{y\mbox{\scriptsize-edge},x}^{(13)}=0$.
}
\label{f:edge_vd}%-----------------------------------------------
\end{center}
\end{figure}

%\csout{We consider the system under open boundary conditions along the $x$ direction.}
Suppose that we have a system with $N_x$ unit cells (each with four sites) in the $x$ direction under the open boundary condition, whereas 
the periodic boundary condition is imposed in the $y$ direction, as illustrated in Fig. \ref{f:open_x}.
Then, we have a single-particle Hamiltonian ${\cal H}(k_y)$
which is a $4N_x\times 4N_x$ matrix.
Thus, the system has $4N_x$ bands as the functions of $k_y$,
\begin{alignat}1
{\cal H}(k_y)\Psi_{n}(k_y)=E_n(k_y)\Psi_{n}(k_y).
\label{EdgHam}%---
\end{alignat}
We show in Figs. \ref{f:edge_qp} and \ref{f:edge_vd}, the spectra of such systems.
These tell that all the spectra of finite size systems are gapped: We cannot find any edge states at or across the zero energy.
However, in Figs. \ref{f:edge_qp}(a) and \ref{f:edge_qp}(d) and in Fig. \ref{f:edge_vd}(d), gapped edgelike states denoted by black points 
which are separated form the bulk can be observed, corresponding to the nontrivial polarization ${\cal P}^{(13)}_x=\pi$ or
${\cal P}^{(14)}_y=\pi$.
These are doubly degenerate bands localized at the left and right boundaries.
The polarization of these states is trivial because of the degeneracy.

\subsection{Entanglement edge state}
\label{s:EntES}%---

If the eP is nontrivial, we can observe the edge states in the corresponding subsystem. 
Let $\Psi_{an}(k_y)$ ($a,n=1,\cdots,4N_x$) be the $n$th eigenfunction of the system 
in Eq. (\ref{EdgHam}) in order of increasing energy.
%Sec. \ref{s:GappedEdge}. 
Then, we can define the projection operator to the 
negative energy states, similarly to Eq. (\ref{ProOpe}),
\begin{alignat}1
P_{-,ab}(k_y)=\sum_{n=1}^{2N_x}\Psi_{an}(k_y)\Psi_{nb}^\dagger(k_y) .
\end{alignat}
In the same way in Sec. \ref{s:EntPol}, we choose two particular sites $(i,j)$ among the four sites in a unit cell.
Then, we have the reduced projection operator:
\begin{alignat}1
P^{(ij)}_{-,ab}(k_y)=P_{-,ab}(k_y), \quad (a,b)=(i,j) \mbox{ mod } 4,
\end{alignat}  
which is a $2N_x\times 2N_x$ matrix.
Let $\xi_n^{(ij)}(k_y)$ ($n=1,\cdots 2N_x$) be the eigenvalue of $P_-^{(ij)}(k_y)$ and let 
$\psi^{(ij)}_n(k_y)$ be the corresponding eigenstate,  similar to Eq. (\ref{EntProOpe}).
Namely, 
\begin{alignat}1
P^{(ij)}_-(k_y)\psi^{(ij)}_n(k_y)=\xi_n^{(ij)}(k_y)\psi^{(ij)}_n(k_y),
\end{alignat}

In Fig. \ref{f:edge_qp}(b), we show the eS $\xi^{(13)}(k_y)$. 
We find the (doubly degenerate) $\xi^{(13)}=1/2$ states, which are zero entanglement energy $\varepsilon^{(13)}=0$ states,
in the spectrum $\xi^{(13)}(k_y)$, indicated by green lines.
These states are localized states at the left end (3) and at the right end  (1) in Fig. \ref{f:open_x}.
Although not shown, the spectrum $\xi^{(24)}(k_y)$ is the same, and we also see the edge states localized 
at the left end (2) and at the right end (4).
Thus, we establish the existence of the left edge state localized at (2) and (3) and of the right edge state localized at (1) and (4).

On one hand, these edge states are due to the bulk-edge correspondence in the eH, namely, 
due to the nontrivial bulk polarization ${\cal P}_x^{(13)}=\pi$.
On the other hand, 
the appearance of the edge states is not only due to the boundaries as in Fig. \ref{f:open_x}, but also due to
the disentanglement of the wave function. 
Namely, the right edge states localized at sites (1) and (4) are strongly entangled  with each other, 
which causes a gap for  the right edge states, as in Fig. \ref{f:edge_qp}(a).
Therefore, if the right edge states localized at sites (14) are disentangled between sites (1) and (4), 
by tracing out (1) or (4), 
states with $\xi\sim1/2$ are expected.
The left edge states localized at sites (2) and (3) are likewise.
Such edge states that appear in the eS will be referred to as the entanglement edge states (eES).

\subsection{Entanglement edge state polarization}
\label{s:EntESP}%---

Next, we switch to the eS, $\xi^{(14)}(k_y)$, and investigate the nature of the eES along the $y$ direction. 
In the partition of $(14)$ and $(23)$ in Fig. \ref{f:open_x}, the right eES belongs to sites $(14)$
and the left eES belongs to sites $(23)$. Therefore, we expect in $\xi^{(14)}$ a fully occupied state 
corresponding to the right eES and a fully unoccupied state corresponding to the left eES.
Indeed, in  Fig. \ref{f:edge_qp}(c),
we see a single completely occupied $\xi_{2N_x}^{(14)}=1$ state
and a single completely unoccupied $\xi_1^{(14)}=0$ state which are clearly isolated from others.
Thus, we have established the eES originated from the edge states localized at the boundaries in the $x$ directions.
However, we have seen that the edge states are gapped in the full spectrum of the original Hamiltonian in Fig. \ref{f:edge_qp} (a).
As we will show below,
it is because these states form the 1D SSH state
propagating to the $y$ direction under periodic boundary condition which should be an insulating gapped state.

To investigate the feature of these gapped edge states, it is useful to calculate the eP.
Namely, for these eES, we define the eP in a similar way in Eqs. (\ref{EntWilLoo}) and (\ref{EntPol}), 
which may be referred to as entanglement edge state polarization (eESP).
Let us define first the entanglement U(1) link variable, 
${\cal U}^{(ij)}_y(k_y)=\psi^{(ij)\dagger}_m(k_y)\psi^{(ij)}_m(k_y+\Delta k)$ for a particular edge mode $m$,
($m=1$ for the fully unoccupied state and $m=2N_x$ for the fully occupied state)
where $\Delta k=2\pi/N$, and,  second, the eWL, 
\begin{alignat}1
{\cal W}^{(ij)}_{x\mbox{\scriptsize -edge},y}
=\prod_{n=1}^N {\cal U}^{(ij)}_y(k_y+n \Delta k).
\end{alignat}
Then, we naturally reach,
\begin{alignat}1
{\cal P}_{x\mbox{\scriptsize-edge},y}^{(ij)}=\arg {\cal W}^{(ij)}_{x\mbox{\scriptsize-edge},y}.
\end{alignat}
%\csout{${\cal P}_{y\mbox{\scriptsize-edge},x}^{(ij)}$ can be similarly defined.}
In the quadrupole phase in Fig. \ref{f:edge_qp}, it turns out that the eESP, ${\cal P}_{x\mbox{\scriptsize-edge},y}^{(14)}=\pi$ both for 
$\xi_{2N_x}^{(14)}=1$ and $\xi_1^{(14)}=0$ states, which are indicated by the red lines in Fig. \ref{f:edge_qp}(c).

So far, we have shown that the system with boundaries in the $x$ direction has gapped edge states with nontrivial 
eESP ${\cal P}_{x\mbox{\scriptsize-edge},y}^{(ij)}=\pi$. This implies that these edge states are themselves topological 1D SSH states
along the boundaries.
Thus, if the system is further cut in the $y$ direction, the 0D zero energy edge states appears. 
These are the BBH corner states discovered in Refs. \cite{Benalcazar61,Benalcazar:2017aa}.

It may be needless to say that
when the system has $N_y$ unit cells along $y$ direction under the open boundary condition, 
we observe similar bulk-edge correspondence, as shown in Fig. \ref{f:edge_qp}(d), \ref{f:edge_qp}(e), and \ref{f:edge_qp}(f).

Next, let us consider the trivial $(0,\pi)$  phase in Fig. \ref{f:edge_vd}.
%, which is specified by bulk $(0,\pi)$ polarization. 
Since this phase is trivial toward the $x$ direction, no zero energy edge states are observed in Fig. \ref{f:edge_vd}(b).
Correspondingly,  no clearly isolated $\xi=0,1$ states are observed.
On the other hand, since the system is nontrivial toward the $y$ direction, zero energy edge states and corresponding $\xi=0,1$ states
are observed in Figs. \ref{f:edge_vd}(e) and \ref{f:edge_vd}(f).
However, such edge states have eESP toward $x$ direction, ${\cal P}_{y\mbox{\scriptsize-edge},x}^{(13)}=0$.
Thus, the eES along the $y$ direction is the trivial SSH state, and hence, we conclude no corner states in this case.

\section{Generic BBH model}
\label{s:GenBBH}%---

The minimal model studied so far is simple enough to obtain the exact bulk eP in Sec. \ref{s:Exact}.
In this section, we will examine a more generic model by including nearest- and next-nearest neighbor hoppings among unit cells,
which are summarized in Appendix \ref{s:App}.
In particular, we introduce several terms with broken time reversal, particle-hole, and chiral symmetries.
For simplicity, we set all parameters zero except for $v$ and $t_{xy}^{\rm i}=t_{yx}^{\rm i}$.
Although we set $\gamma_x=\gamma_y\equiv \gamma$ and $\lambda_x=\lambda_y\equiv\lambda$,
$C_4$ symmetry is broken by the $v$ term.
We will show that the topological quadrupole phase
is characterized by the bulk eP and eESP even for such a generic model.

\begin{figure}[htb]
\begin{center}
\begin{tabular}{cc}
\includegraphics[scale=0.33]{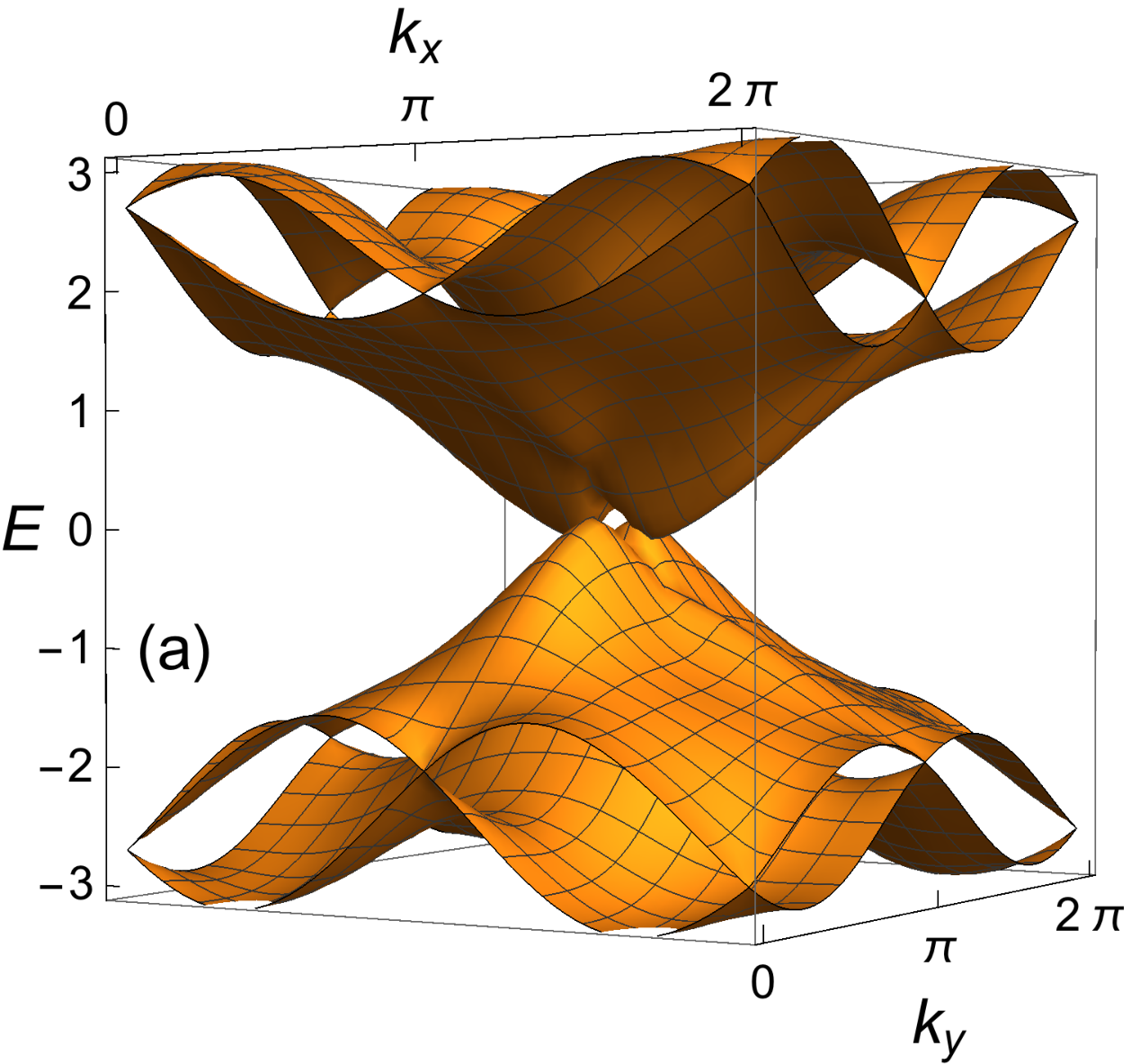}
&
\includegraphics[scale=0.33]{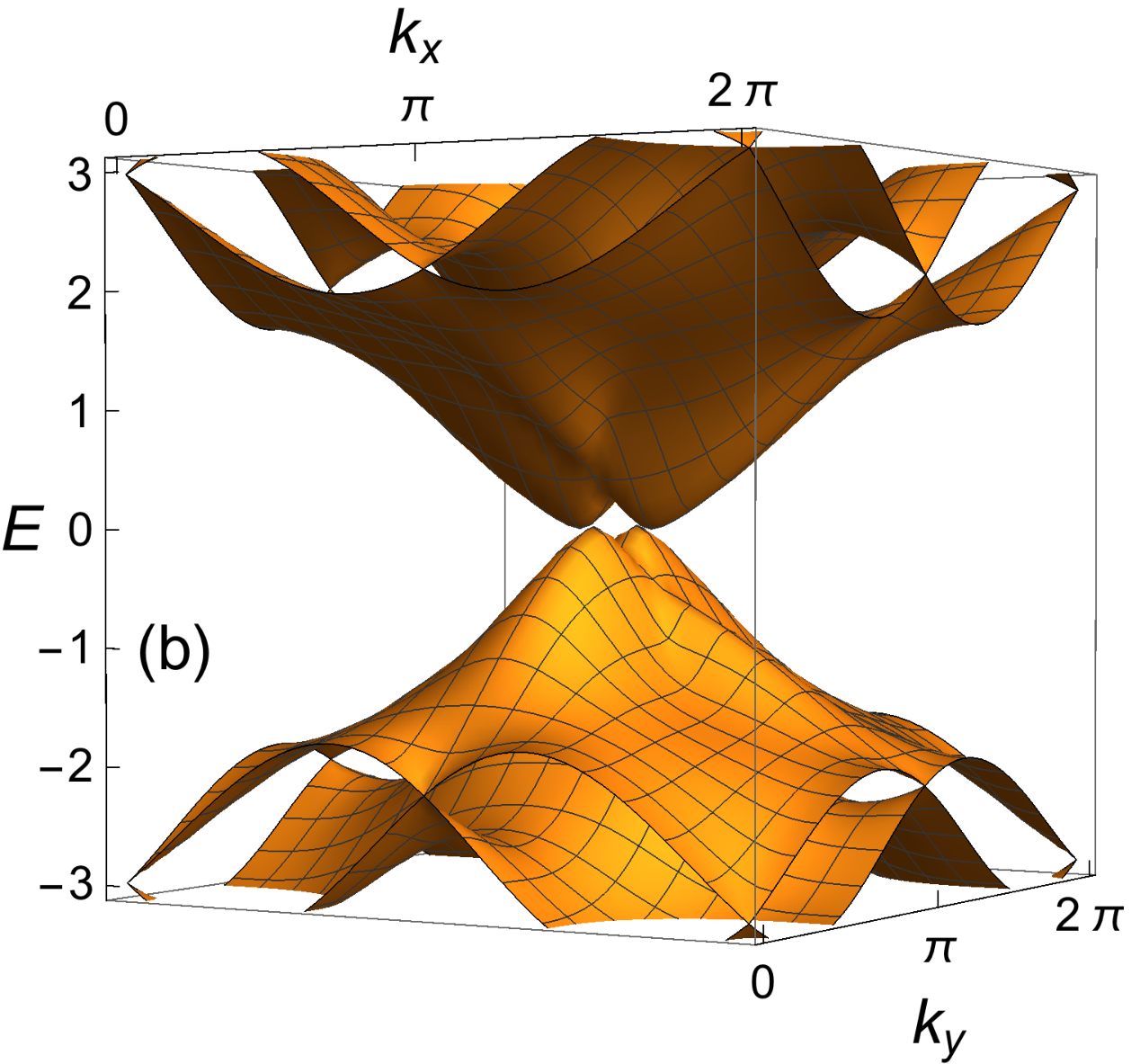}
\end{tabular}
\caption{Spectra of generic BBH model with $\lambda=1$, and $v=t_{xy}^{\rm i}=t_{yx}^{\rm i}=0.3$.
(a)  belongs to topological phase ($\gamma=0.9$), whereas (b) belongs to the trivial phase ($\gamma=1.1$).
Even with finite $v$ and $t_{xy}^{\rm i}=t_{yx}^{\rm i}$, the degeneracies at the high-symmetry points
$(k_x^*,k_y^*)$ imply the reflection symmetries Eqs. (\ref{RefSym}).
}
\label{f:ssh_nnn}%-----------------------------------------------
\end{center}
\end{figure}

In Fig. \ref{f:ssh_nnn}, we show the spectra of the model near the topological-trivial transition point ($\gamma=1$).
These are semimetalic, but the finite direct gap enables us to define the bulk topological invariant.
The double-degeneracy of the spectrum in the minimal model is lifted except for the high-symmetry points. 
The degeneracies at these points are due to the 
reflection symmetries associated with $M_x$ and $M_y$ \cite{Benalcazar:2017aa}.
Thus, for the present model, the projection operator to the ground state Eq. (\ref{ProOpe}) cannot be given by Eq. (\ref{ProOpeMin}).

\begin{figure}[htb]
\begin{center}
\begin{tabular}{cc}
\includegraphics[scale=0.33]{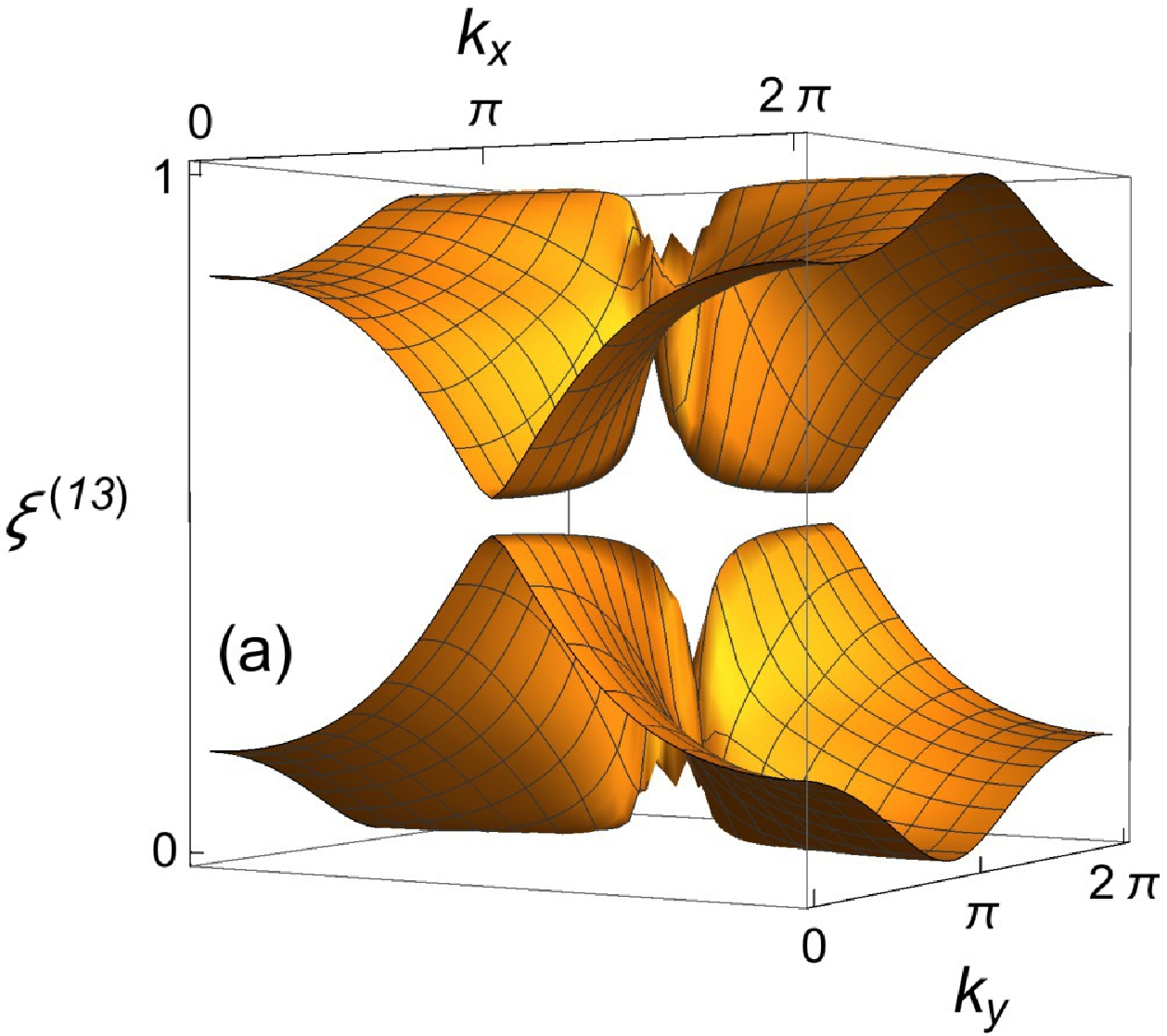}
&
\includegraphics[scale=0.33]{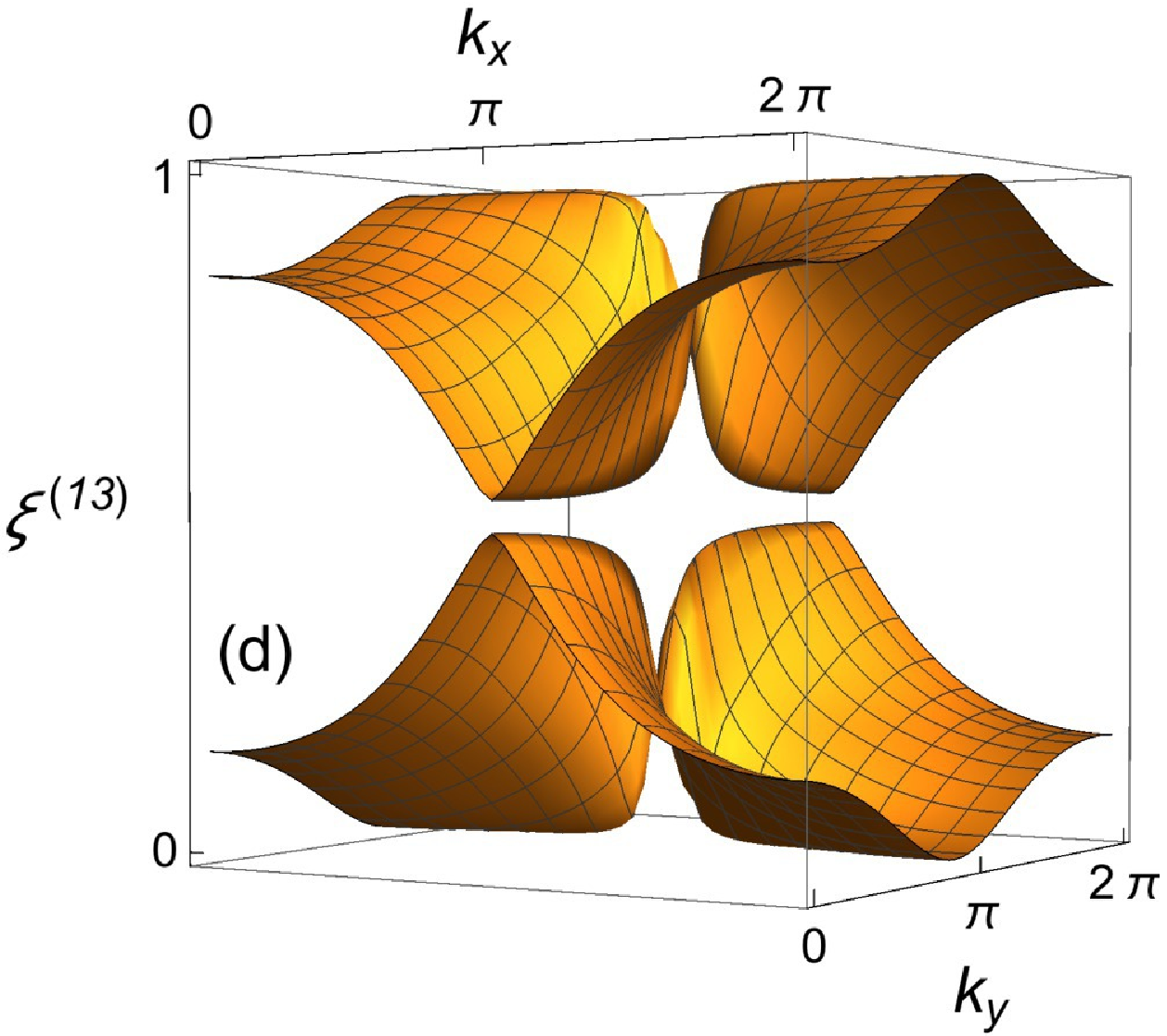}
\\
\includegraphics[scale=0.33]{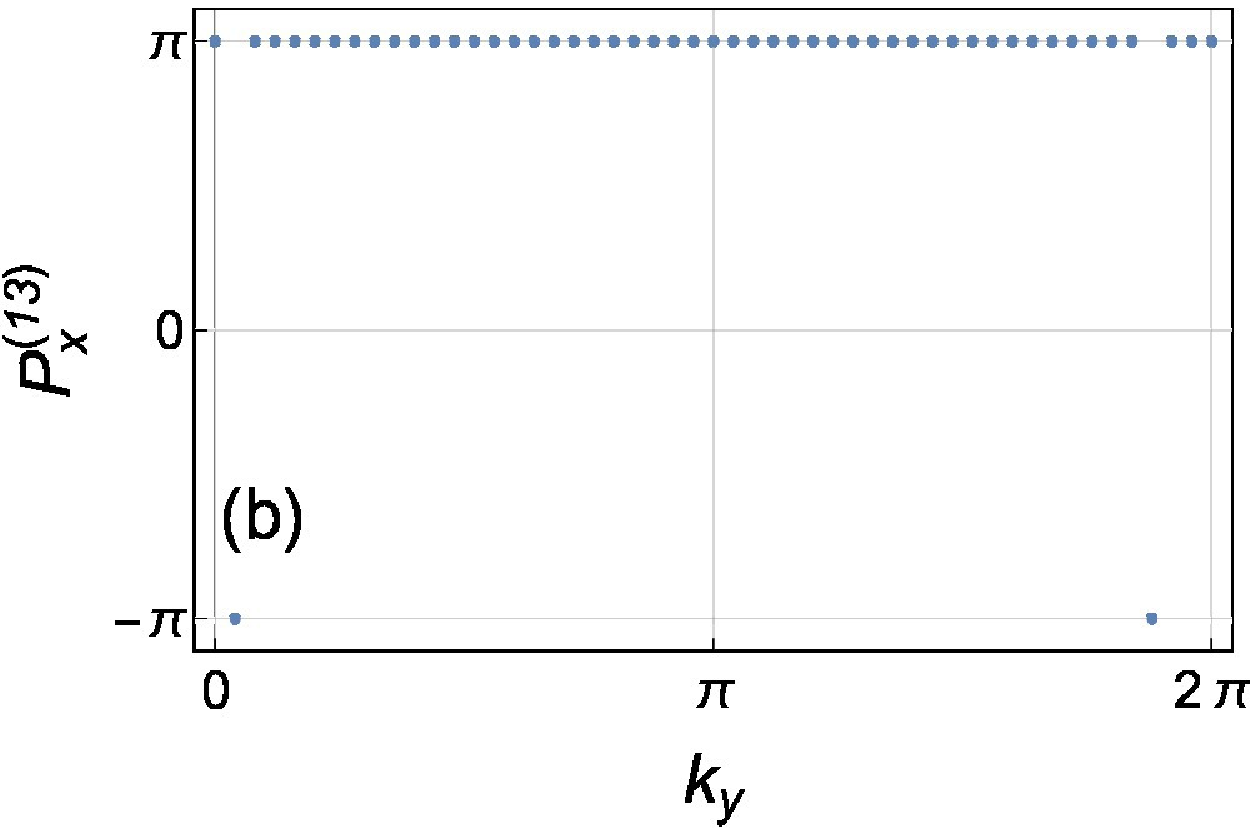}
&
\includegraphics[scale=0.33]{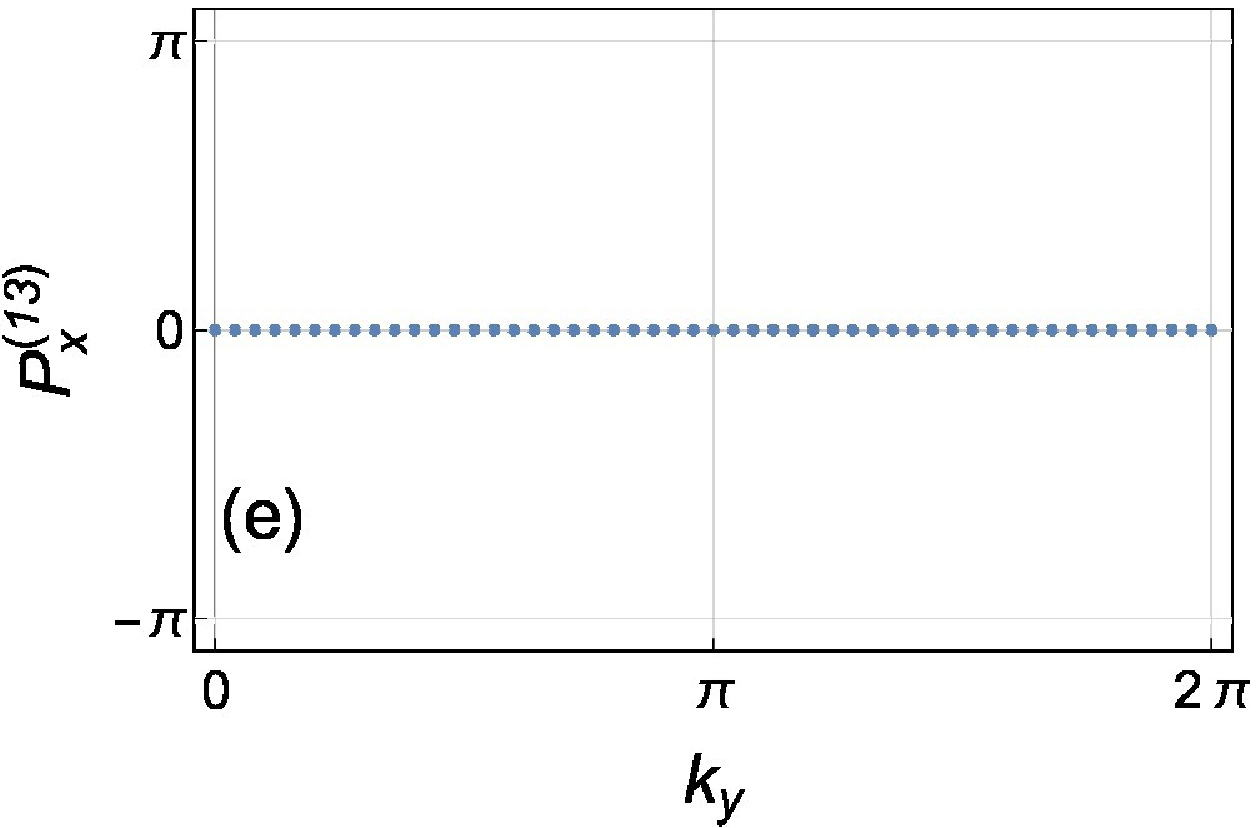}
\\
\includegraphics[scale=0.33]{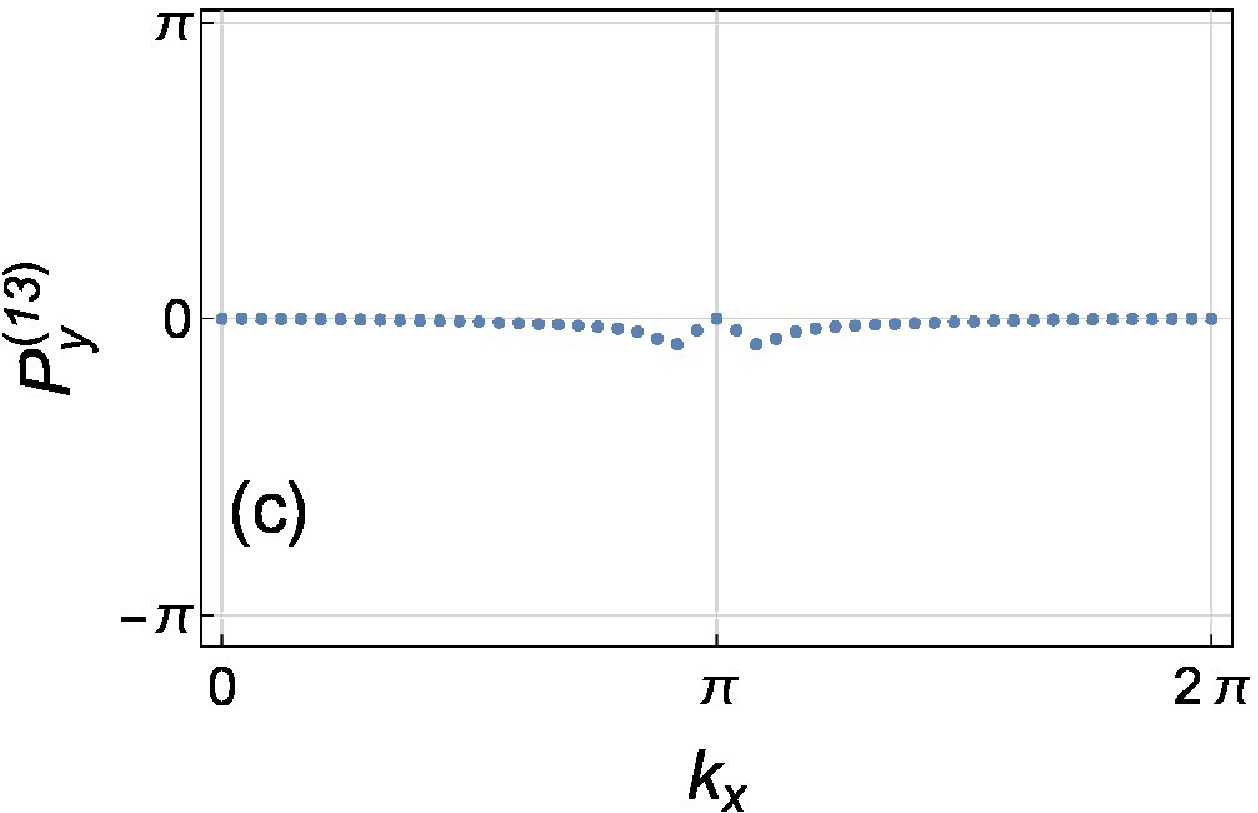}
&
\includegraphics[scale=0.33]{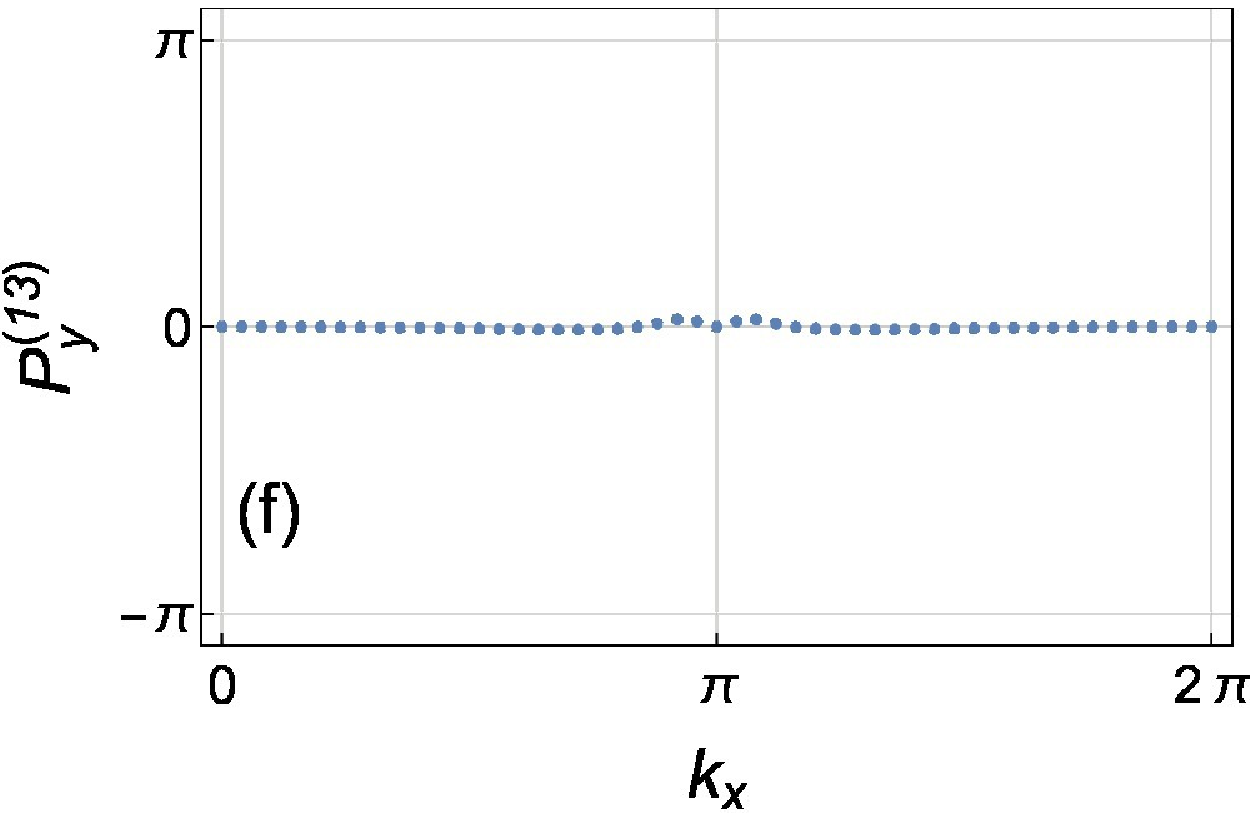}
\end{tabular}
\caption{
(a), (d)  eS $\xi^{(13)}(k)$. 
(b), (e)  eP ${\cal P}^{(13)}_x(k_y)$.
(c), (f)  eP ${\cal P}^{(13)}_y(k_x)$.
(a), (b), (c) are in the topological quadrupole phase corresponding to (a) in Fig. \ref{f:ssh_nnn}, and 
(d), (e), (f) are in the trivial phase corresponding to (b) in Fig. \ref{f:ssh_nnn}.
%(d)  eS $\xi^{(13)}(k)$,  
%(e)  eP ${\cal P}^{(13)}_x(k_y)$, 
%and (f) eP ${\cal P}^{(13)}_y(k_x)$ for (b) in Fig. \ref{f:ssh_nnn}.
%Left and Right figures correspond to the model in Fig. \ref{f:ssh_nnn} (a) and (b), respectively.
%(a) and (d) show the eS $\xi^{(13)}(k)$, 
%(b), (e) and (c), (f) show the corresponding eP ${\cal P}^{(13)}_x(k_y)$ and ${\cal P}^{(13)}_y(k_x)$, respectively.
}
\label{f:nnn_bulk}%-----------------------------------------------
\end{center}
\end{figure}

For such generic model, let us compute the bulk topological invariant defined by Eq. (\ref{EntPol}).
In Fig. \ref{f:nnn_bulk}, we show the eS $\xi^{13}(k)$. 
The occupied and empty bands are indeed gapped at $\xi=1/2$. The gap is small since the model is near the
transition point.
Using the wave function of the occupied band above, we can compute the eWL and eP. 
The eP ${\cal P}_x^{(13)}(k_y)$ is exactly quantized as $\pi$ in the topological phase and $0$ in the trivial phase,
as shown in Figs. \ref{f:nnn_bulk}(b) and \ref{f:nnn_bulk}(e), respectively. 
The quantization is due to the reflection symmetry $M_x$, Eq. (\ref{RefSymEnt}).
On the other hand, the eP ${\cal P}_y^{(13)}(k_x)$ is slightly fluctuating as  the  function of $k_x$, 
as can be seen in Figs. \ref{f:nnn_bulk}(c) and \ref{f:nnn_bulk}(f), 
since reflection symmetries give no constraints on it.
Although not shown in the figure, we have  ${\cal P}_y^{(14)}(k_y)=\pi$ in the topological phase and $=0$ in the trivial phase,
so that we have the bulk topological invariant,
$({\cal P}_y^{(13)},{\cal P}_y^{(14)})=(\pi,\pi)$ in the topological phase and $(0,0)$ in the trivial phase.
\begin{table}[htb]
\begin{tabular}{c|c|c|c|c|c}
Phase &Ref. ev & $(0,0)$ & $(\pi,0)$ &$(0,\pi)$ & $(\pi,\pi)$\\
\hline\hline
QP & $p^{(13)}_{k^*}$ &$-1$& $+1$ &$-1$&$+1$\\
& $p^{(14)}_{k^*}$ &$-1$& $-1$ &$+1$&$$+1\\
\hline
Tr & $p^{(13)}_{k^*}$ &$-1$& $-1$ &$-1$&$-1$\\
& $p^{(14)}_{k^*}$ &$-1$& $-1$ &$-1$&$-1$\\
\end{tabular}
\caption{
Eigenvalues  (ev) of the reflection operators at the high symmetry points. 
The phase QP and Tr correspond to the quadrupole and trivial phase in Fig. \ref{f:ssh_nnn} (a) and (b), respectively. 
}
\label{t:Ref}
\end{table}
We also computed the eigenvalues of the reflection operators $M_x$ and $M_y$ at the high symmetry points.
In Table \ref{t:Ref}, we find that the relation Eq. (\ref{RefEig}) holds for the present generic model.

\begin{figure}[htb]
\begin{center}
\begin{tabular}{cc}
\includegraphics[scale=0.33]{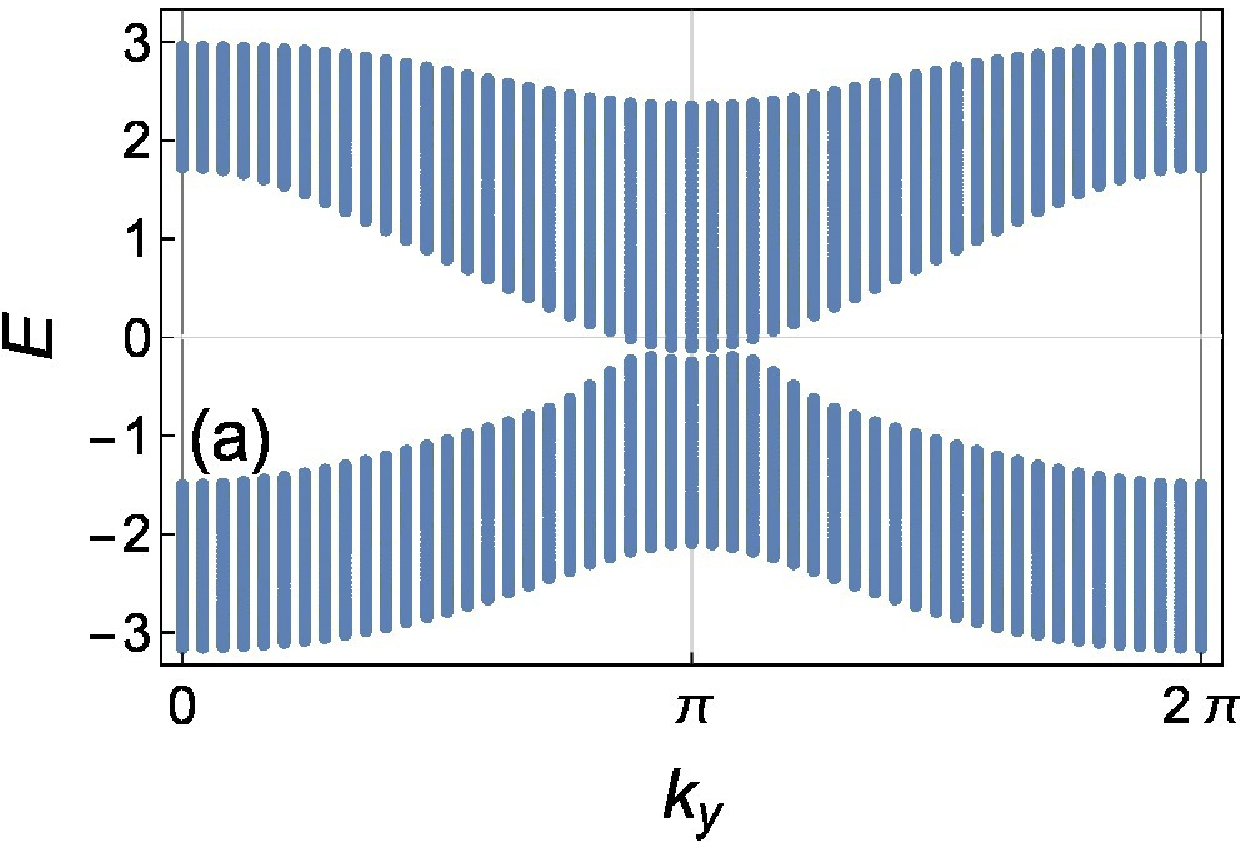}
&
\includegraphics[scale=0.33]{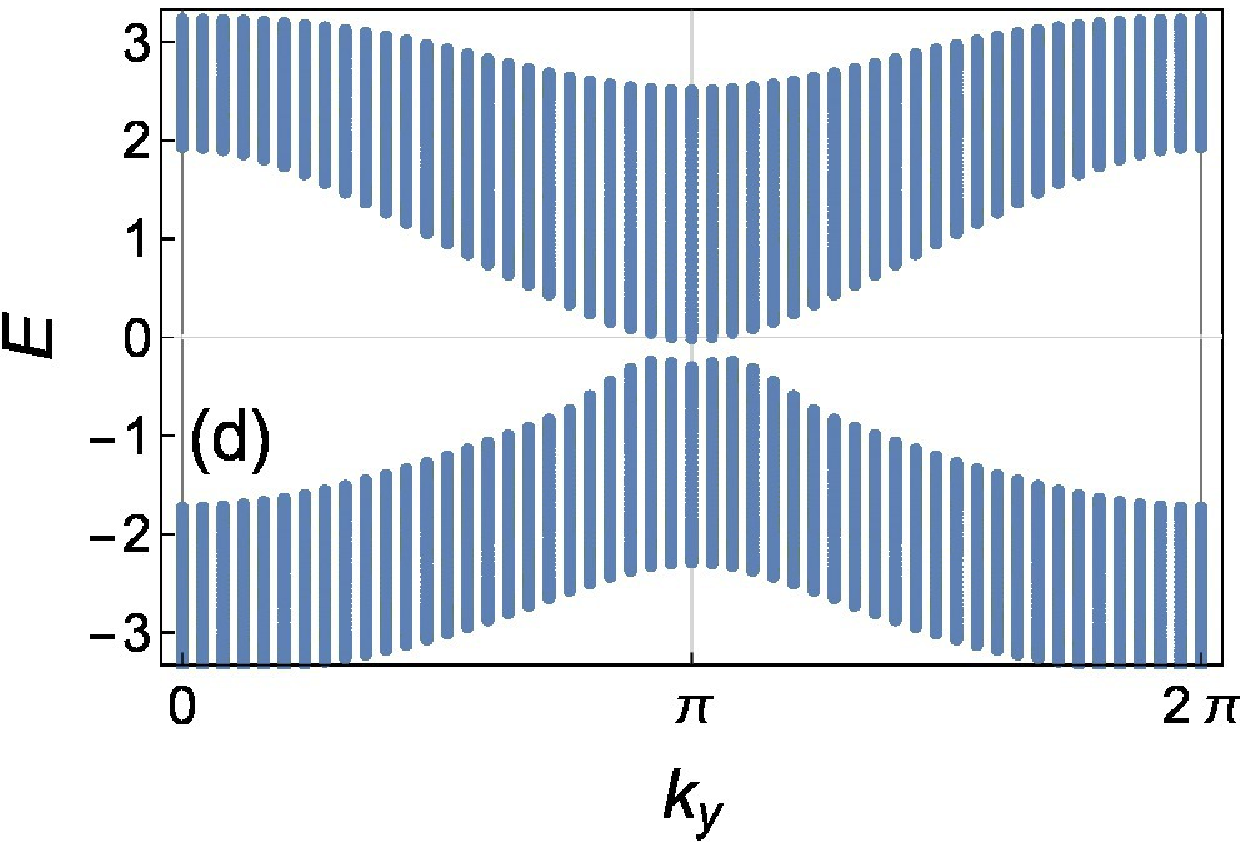}
\\
\includegraphics[scale=0.33]{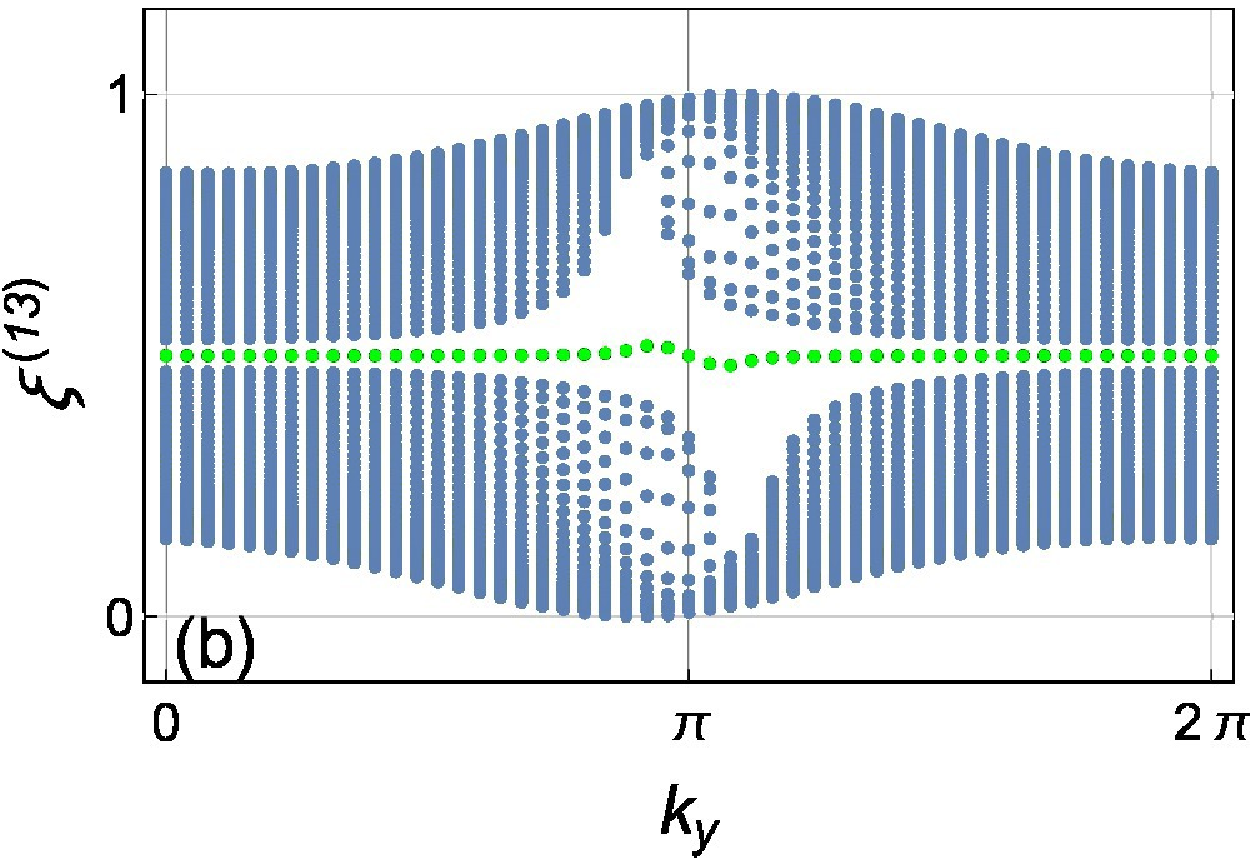}
&
\includegraphics[scale=0.33]{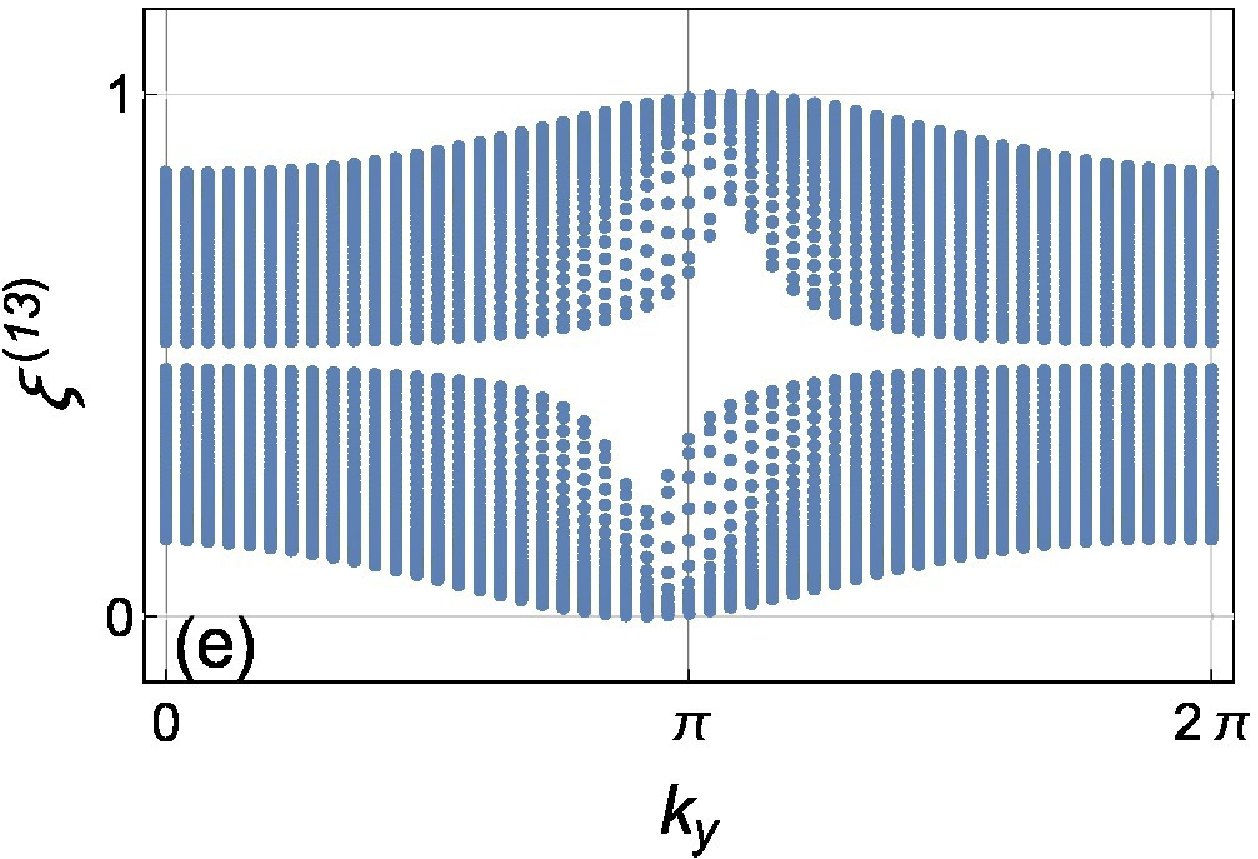}
\\
\includegraphics[scale=0.33]{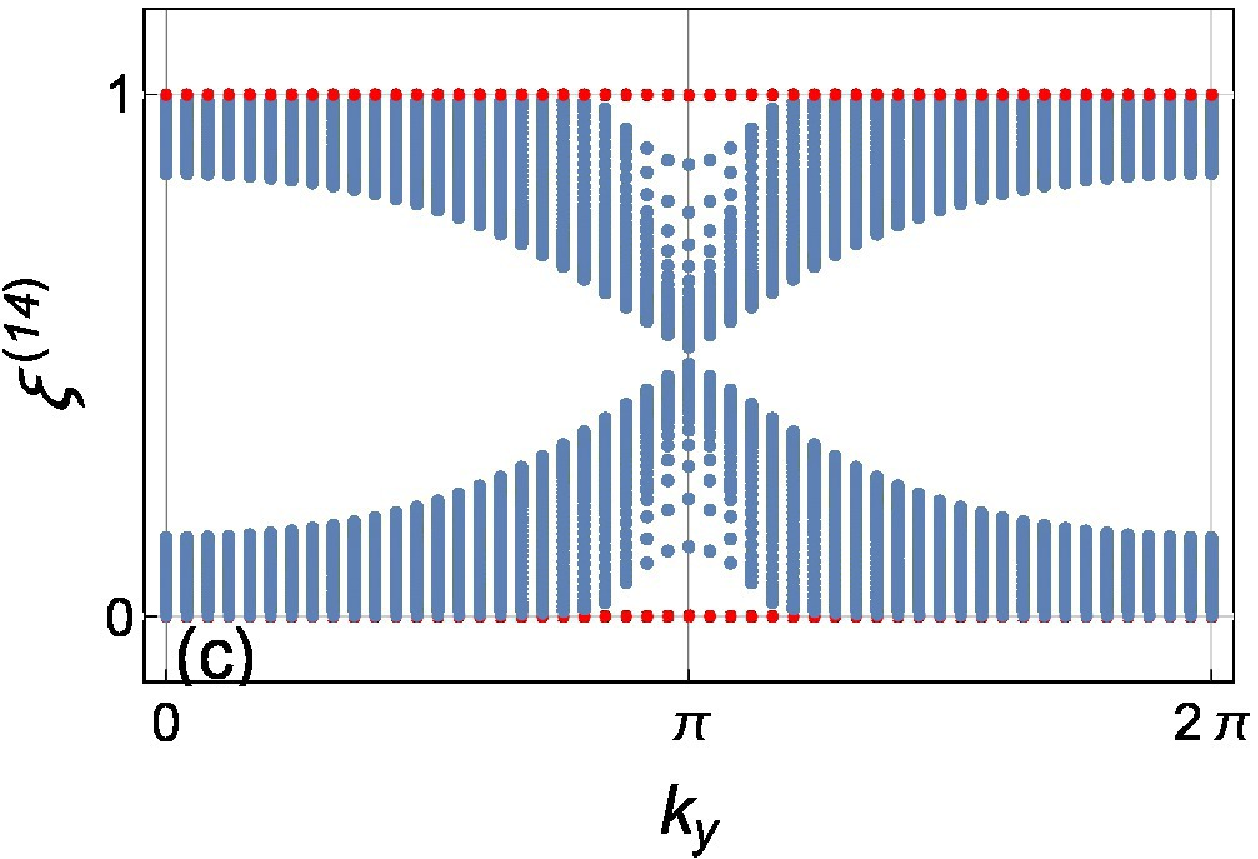}
&
\includegraphics[scale=0.33]{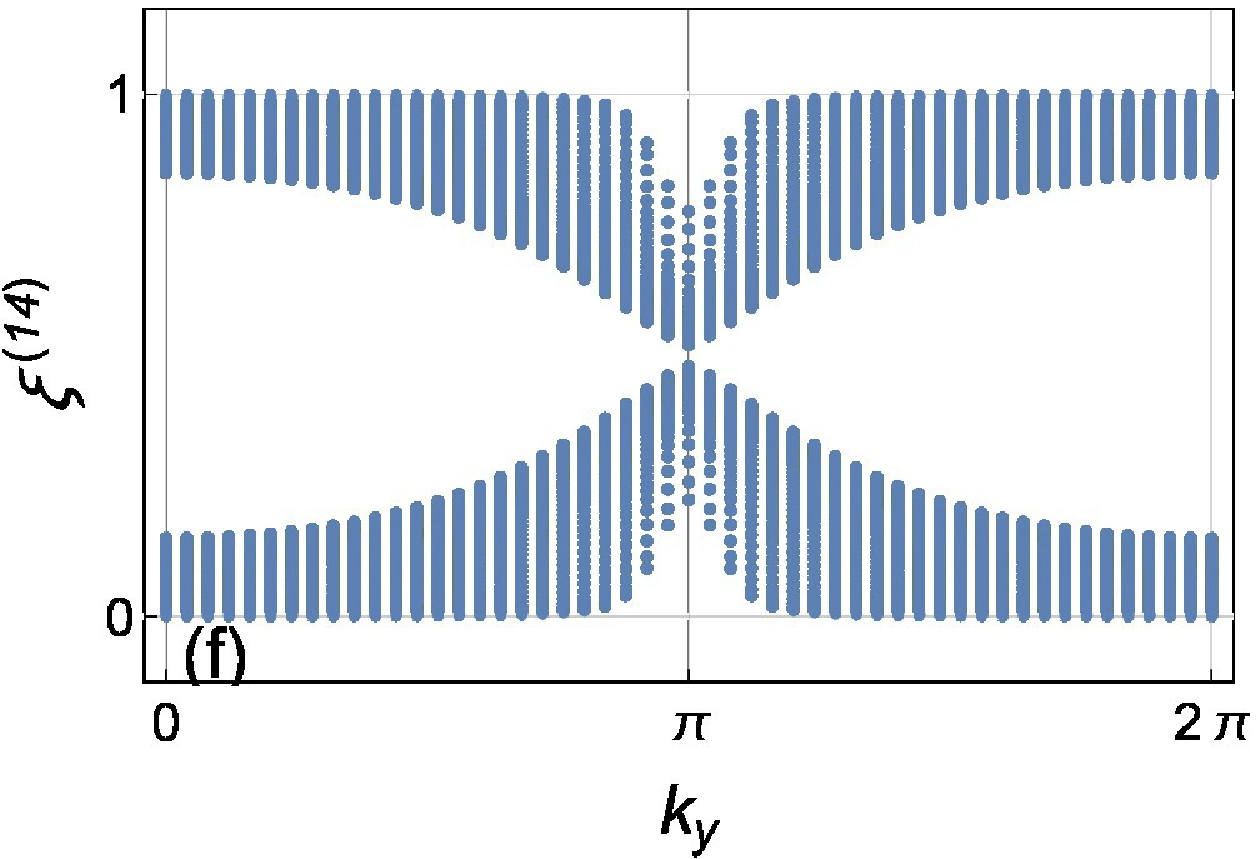}
\end{tabular}
\caption{
(a), (d) Spectra of the Hamiltonian with boundaries.
(b), (e)  eS $\xi^{(13)}(k_y)$. (c), (f) eS $\xi^{(14)}(k_y)$. 
%for (a) in Fig. \ref{f:ssh_nnn}.
(a), (b), (c) are in the topological quadrupole phase corresponding to (a) in Fig. \ref{f:ssh_nnn}, and 
(d), (e), (f) are in the trivial phase corresponding to (b) in Fig. \ref{f:ssh_nnn}.
%(d) Spectrum of the Hamiltonian with boundaries,
%(e) corresponding eS $\xi^{(13)}(k_y)$, (f)  corresponding $\xi^{(14)}(k_y)$ for (b) in Fig. \ref{f:ssh_nnn}:
%Spectra of the system with boundaries in the $x$ direction.
%Left and right correspond to Fig. \ref{f:ssh_nnn} (a) and (b), respectively. 
%(a) and (d) show the spectra of the Hamiltonian with boundaries.
%(b)  and (e), and (c) and (f) show the corresponding eS $\xi^{(13)}(k_y)$ and $\xi^{(14)}(k_y)$,
%respectively.
}
\label{f:nnn_edge}%-----------------------------------------------
\end{center}
\end{figure}

In Fig. \ref{f:nnn_edge}, we show various spectra for the system with boundaries in the $x$ direction.
First, we would like to mention the characteristic property of the spectra.
It is noted that the spectra of the Hamiltonian, Figs. \ref{f:nnn_edge}(a) and \ref{f:nnn_edge}(d) are asymmetric with respect to $E=0$.
This implies broken chiral and particle-hole symmetries.
Also in the eS, we can observe broken symmetries:
When the system has chiral or time reversal symmetry, the eS obeys $\xi^{(13)}_n(k_y)=\xi^{(13)}_n(-k_y)$
for systems with boundaries, which can be easily 
derived from the bulk properties in Eq. (\ref{ChiAddSym}) or (\ref{TimAddSym}).
Indeed, in Figs. \ref{f:edge_qp} and \ref{f:edge_vd}, we observe such spectra as are symmetric with respect to $k_y=\pi$. 
Only with reflection symmetries, however, the most generic spectrum has the property Eq. (\ref{RefSpeSym}),
$\xi^{(13)}_n(k_y)+\xi^{(13)}_{2N_x+1-n}(-k_y)=1$.
Thus, the eS, Figs. \ref{f:nnn_edge}(b) and \ref{f:nnn_edge}(e), tell broken chiral and time reversal symmetries for the present model.
On the other hand, symmetry of $\xi^{(14)}_n(k_y)=\xi^{(14)}_{n}(-k_y)$ is ensured only by the reflection 
with respect to $M_y$ denoted by Eq. (\ref{RefSymCon}).
Thus, we find the eS, Figs. \ref{f:nnn_edge}(c) and \ref{f:nnn_edge}(f),  symmetric with respect to $k_y=\pi$.

Next, let us switch to the topological property of the model.
In Fig. \ref{f:nnn_edge} (b), almost zero entanglement energy states denoted by green points 
which are isolated from other bulk bands are observed. 
This is the doubly degenerate edge state localized at (3) of the left edge and at (1) of the right edge in Fig. \ref{f:open_x}.
Because of the remaining entanglement due to the introduced generic hopping terms, the zero energy is slightly lifted.
Nevertheless, they are clearly separated from the bulk spectrum.
In Fig. \ref{f:nnn_edge} (c),  we find that the degenerated zero energy  edge states at (1) and (3) in Fig. \ref{f:nnn_edge} (b)
is completely lifted in the spectrum of $\xi^{(14)}$: Namely, the edge state localized 
at the right (14)-sites appears as a fully occupied band, whereas the other edge state localized at the left (23)-sites becomes 
a fully unoccupied band, as denoted by red points in the figure.
Since the model is near the transition point, the gap between the edge states and the bulk states seems very small.
Nevertheless, there is enough gap to separate in the practical numerical computations.
Then, we can compute the eESP for these edge states. The result is  ${\cal P}_{x\mbox{\scriptsize-edge},y}^{(14)}=\pi$ for
both bands. 
Thus, it turns out that the edge states along the boundaries are basically 1D topological SSH states. 
This implies that if the present system, which is finite in the $x$ direction, becomes also finite in the $y$ direction, 
the SSH edge states show the 0D zero energy edge states. These are observed by the corner states.

On the other hand, in the trivial phase, no edge states are observed in Figs. \ref{f:nnn_edge}(e) and \ref{f:nnn_edge}(f).
Thus, this system is also trivial even in the sense of the second-order topological insulator.

\section{Summary and discussion}
\label{s:SumDis}%---

In this paper, we formulated the eP to describe the topological quadrupole phase.
As the conventional polarization is defined by the Berry or Zak phase of the occupied states of the Hamiltonian $H$,
the eP is also that of the occupied states of the {\it eH} ${\cal H}$ and $\bar{\cal H}$.
More generically, we expect that some classes of symmetry protected topological phase can be easily described by the topological 
properties not only of $H$ but also of $({\cal H},\bar{\cal H})$.
This method is simple enough to obtain the exact eP for the minimal BBH model, which is consistent to the known results.
Furthermore, the eP reveals the topological property of the edge states: They are themselves the 1D SSH states, and thus, 
if they have boundaries, they can show the 0D edge states.
These are the corner states for a system under the open boundary condition both for the $x$ and $y$ directions. 
This feature can be directly checked by the eESP.

The adiabatic dipole pump proposed in \cite{Benalcazar61}
is also revealed by the eP.  Let us consider the minimal BBH model Eqs. (\ref{Ham}) and (\ref{HamEle}),
and assume that $\delta$, $\lambda_{x,y}$, and $\gamma_{x,y}$ depend on time $t\in[0,2\pi]$.  
In the process $t:0\rightarrow \pi$ of 
$(\delta(t),\lambda_{x,y}(t),\gamma_{x,y})=(\delta^0\cos t,\lambda^0_{x,y}\sin t,0)$,
the eP is changed by the amount of the Chern numbers of the simple $2\times 2$ ``Hamiltonians" $P_-^{(13)}(k)$ and $P_-^{(14)}(k)$,
$(c^{(13)},c^{(14)})=(-1,-1)$, and their complements $(c^{(24)},c^{(23)})=(1,1)$, 
which were referred to as the entanglement Chern numbers (eCN) \cite{Fukui:2014qv}.
The pattern of eCN above matches the dipole pump arising from the bulk quadrupole moment 
with vanishing bulk dipole moment, as discussed in Ref. \cite{Benalcazar61}.
In the other process $t:\pi \rightarrow 2\pi$ of $(\delta(t),\lambda_{x,y}(t),\gamma_{x,y})=(\delta^0\cos t,0,-\gamma^0_{x,y}\sin t)$,
we readily find the vanishing eCN.
%\csout{Thus, in the quadrupole phase, the eCN in the adiabatic pumping process is given by $(-1,-1)$.}
Even for the more generic model derived in Appendix \ref{s:App}, 
it is easy to compute such entanglement Chern numbers by the method proposed in  Ref. \cite{FHS05}.
Detailed analysis, including more generic and realistic models, will be published elsewhere.

The eP presented in this paper is also applicable to the topological octupole phase, which will be published in future.

%The present technique can be easily applied to the topological octupole phase.

\section*{Acknowledgments}
We would like to thank K.-I. Imura and Y. Yoshimura for fruitful discussions.
This work was supported in part by Grants-in-Aid for Scientific Research No. 17K05563, No. 17H06138, 
and No. 16K13845 from
the Japan Society for the Promotion of Science.

\appendix

%\begin{widetext}

\section{Generic BBH model with nearest and next nearest neighbor hoppings among unit cells}
\label{s:App}%-------

%\begin{figure}[htb]
%\begin{center}
%\begin{tabular}{c}
%\includegraphics[scale=0.45]{figs/ssh_1.eps}\\
%\includegraphics[scale=0.45]{figs/ssh_2.eps}
%\end{tabular}
%\caption{$u_j$ are real parameters, while $s$ and $t$ are complex parameters 
%$s=s^{\rm r}+is^{\rm i}$ and $t=t^{\rm r}+it^{\rm i}$
%with a constraint $s^{\rm r}=-t^{\rm r}$.
%}
%\label{f:ssh_a}
%\end{center}
%\end{figure}

In this Appendix, we derive most generic Hamiltonian with reflection symmetries Eqs. (\ref{RefSym})
up to next-nearest neighbor hoppings between unit cells.

\begin{table}
\begin{tabular}{l|l|c|c|c||c|c|c}
Categ.&Matrices&Part.&$K$&$C$&$T$&$C$&$\gamma_5$\\
\hline\hline
$R_{\rm ee}$&$i\gamma_2\gamma_4=-\sigma_2\otimes\sigma_3$&12&$-$&$-$&$\times$&&$\times$\\
%\cline{2-4}
& $\gamma_4=\1\otimes\sigma_1$&13&$+$&$-$&&&\\
& $\gamma_2=-\sigma_2\otimes\sigma_2$&14&$+$&$-$&&&\\
\hline
$R_{\rm oo}$&$\gamma_5=\1\otimes\sigma_3$&d&$+$&$+$&&$\times$&$\times$\\
&$i\gamma_1\gamma_3=\sigma_2\otimes\1$&12&$-$&$-$&$\times$&&$\times$\\
&$i\gamma_4\gamma_5=\1\otimes\sigma_2$&13&$-$&$+$&$\times$&$\times$&\\
&$i\gamma_2\gamma_5=\sigma_2\otimes\sigma_1$&14&$-$&$+$&$\times$&$\times$&\\
\hline\hline
$R_{\rm eo}$&$i\gamma_1\gamma_2=-\sigma_3\otimes\1$&d&$+$&$+$&$\times$&&$\times$\\
&$i\gamma_1\gamma_4=-\sigma_1\otimes\sigma_3$&12&$+$&$+$&$\times$&&$\times$\\
&$i\gamma_3\gamma_5=\sigma_3\otimes\sigma_1$&13&$+$&$-$&$\times$&$\times$&\\
&$\gamma_1=-\sigma_1\otimes\sigma_2$&14&$-$&$+$&&&\\
\hline
$R_{\rm oe}$&$i\gamma_3\gamma_4=-\sigma_3\otimes\sigma_3$&d&$+$&$+$&$\times$&&$\times$\\
&$i\gamma_2\gamma_3=-\sigma_1\otimes\1$&12&$+$&$+$&$\times$&&$\times$\\
&$\gamma_3=-\sigma_3\otimes\sigma_2$&13&$-$&$+$&&&\\
&$i\gamma_1\gamma_5=\sigma_1\otimes\sigma_1$&14&$+$&$-$&$\times$&$\times$&
\end{tabular}
\caption{
Classification of the additional Hamiltonian terms with reflection symmetries Eqs. (\ref{RefSym}).
The sign $\pm$ denotes $S\gamma S^{-1}=\pm S$, where $S=K,C=\gamma_5K$.
The symbol $\times$ stands for the symmetry breaking for $T$, $C$ and chiral ($\gamma_5$) symmetries.
}
\label{t:gham}
\end{table}

Firstly, diagonal hopping terms are
\begin{alignat}1
H_{\rm d}=\sum_j\big[&iu_x
(c_{1j}^\dagger c_{1j+\hat x}-c_{2j}^\dagger c_{2j+\hat x}-c_{3j}^\dagger c_{3j+\hat x}+c_{4j}^\dagger c_{4j+\hat x})
\nonumber\\
+&iu_y(c_{1j}^\dagger c_{1j+\hat y}-c_{2j}^\dagger c_{2j+\hat y}+c_{3j}^\dagger c_{3j+\hat y}-c_{4j}^\dagger c_{4j+\hat y})
\nonumber\\
+&tc_{1j}^\dagger c_{1j+\hat x+\hat y}+t^*c_{2j}^\dagger c_{2j+\hat x+\hat y}
\nonumber\\
+&sc_{3j}^\dagger c_{3j+\hat x+\hat y}+s^*c_{4j}^\dagger c_{4j+\hat x+\hat y}
\nonumber\\
+&sc_{1j}^\dagger c_{1j-\hat x+\hat y}+s^*c_{2j}^\dagger c_{2j-\hat x+\hat y}
\nonumber\\
+&tc_{3j}^\dagger c_{3j-\hat x+\hat y}+
t^*c_{4j}^\dagger c_{4j-\hat x+\hat y}\big]
\nonumber\\
+&\mbox{H.c.} ,
\label{Hop_d}%---
\end{alignat}
where $u_x$ and $u_y$ are real parameters while $s=s^{\rm r}+is^{\rm i}$ and $t=t^{\rm r}+it^{\rm i}$ are
complex parameters with the constraint $s^{\rm r}=-t^{\rm r}$.
Similar to Eqs. (\ref{HamMom}) and (\ref{HamEle}), this Hamiltonian can be expressed in the Brillouin zone as
\begin{alignat}1
h_{\rm d}=&2\sin k_x\big[ u_x+(t^{\rm i}-s^{\rm i})\cos k_y\big]i\gamma_3\gamma_4
\nonumber\\
+&2 \big[ u_y +(t^{\rm i}+s^{\rm i})\cos k_x\big]\sin k_y i\gamma_1\gamma_2
\nonumber\\
-&4t^{\rm r}\sin k_x\sin k_y\gamma_5.
\nonumber\\
\end{alignat}
%\begin{figure}[htb]
%\begin{center}
%\includegraphics[scale=0.45]{figs/ssh_3.eps}%\hspace*{5mm}
%\caption{$v$, $v_x$,  $v_y$, and $v_{xy}$ are are real parameters.
%}
%\label{f:ssh_12}
%\end{center}
%\end{figure}
Next, the hopping terms between 1 and 2 and between 3 and 4 are 
\begin{alignat}1
H_{\rm 12}=i\sum_j\big[&
c_{1j}^\dagger(vc_{2j}+v_xc_{2j+\hat x}+v_y c_{2j+\hat y}+v_{xy}c_{2j+\hat x+\hat y})
\nonumber\\
-&c_{3j}^\dagger (vc_{4j}+v_xc_{4j-\hat x}+v_yc_{4j+\hat y}+v_{xy}c_{4j-\hat x+\hat y})\big]
\nonumber\\
+&\mbox{H.c.} , 
\label{Hop_12}%---
\end{alignat}
where $v$, $v_x$, $v_{y}$, and $v_{xy}$ are all real parameters. 
In the momentum space, it gives
\begin{alignat}1
h_{\rm 12}=&(v+v_x \cos k_x+v_y \cos k_y +v_{xy} \cos k_x\cos k_y)i\gamma_2\gamma_4
\nonumber\\
&+\sin k_x(v_x+v_{xy}\cos k_y)i\gamma_2\gamma_3
\nonumber\\
&+(v_y+v_{xy} \cos k_x) \sin k_y i\gamma_1\gamma_4
\nonumber\\
&+v_{xy}\sin k_x\sin k_y i\gamma_1\gamma_3. 
\end{alignat}
%\begin{figure}[htb]
%\begin{center}
%\includegraphics[scale=0.45]{figs/ssh_e.eps}%\hspace*{5mm}
%\caption{$t_j=(t_j^{\rm r}+it_j^{\rm i})$, where $t_j^{\rm r}$ and $t_j^{\rm i}$ are real parameters.
%}
%\label{f:ssh_e}%-----------------------------------------------
%\end{center}
%\end{figure}
Likewise, we obtain the hopping terms between 1 and 3, and between 2 and 4,
\begin{alignat}1
H_{\rm 13}=\sum_j\big[&
c_{1j}^\dagger(t_yc_{3j+\hat y}+t_y^*c_{3j-\hat y}
\nonumber\\
&\hspace{10mm}+t_{yx} c_{3j+\hat x-\hat y}+t_{yx}^*c_{3j+\hat x+\hat y})
\nonumber\\
+&c_{2j}^\dagger (t_yc_{4j-\hat y}+t_y^*c_{4j+\hat y}
\nonumber\\
&\hspace{10mm}+t_{yx}c_{4j-\hat x+\hat y}+t_{yx}^*c_{4j-\hat x-\hat y})\big]
\nonumber\\
+&\mbox{H.c.} ,
\label{Hop_13}%---
\end{alignat}
where $t_j$ ($j=y,yx$) is a complex parameter which we set $t_j=t_j^{\rm r}+it_j^{\rm i}$.  
In the momentum representation, this then reduces to
\begin{alignat}1
h_{\rm 13}=&2(t_y^{\rm r} +t_{yx}^{\rm r}\cos k_x)\cos k_y \gamma_4
\nonumber\\
+&2(-t_y^{\rm i}+t_{yx}^{\rm i}\cos k_x)\sin k_yi\gamma_3\gamma_5
\nonumber\\
+&2t_{yx}^{\rm r}\sin k_x\cos k_y \gamma_3-2t_{yx}^{\rm i}\sin k_x\sin k_y i\gamma_4\gamma_5.
\end{alignat}
%\begin{figure}[htb]
%\begin{center}
%\includegraphics[scale=0.45]{figs/ssh_d.eps}%\hspace*{5mm}
%\caption{$t_j=(t_j^{\rm r}+it_j^{\rm i})$, where $t_j^{\rm r}$ and $t_j^{\rm i}$ are real parameters.
%}
%\label{f:ssh_14}
%\end{center}
%\end{figure}
Finally, the hopping terms between 1 and 4, and between 2 and 3, we find
\begin{alignat}1
H_{\rm 14}=\sum_j\big[&
c_{1j}^\dagger(t_xc_{4j+\hat x}+t_x^*c_{4j-\hat x}
\nonumber\\
&\hspace{10mm}+t_{xy} c_{4j+\hat x+\hat y}+t_{xy}^*c_{4j-\hat x+\hat y})
\nonumber\\
-&c_{2j}^\dagger (t_xc_{3j-\hat x}+t_x^*c_{3j+\hat x}
\nonumber\\
&\hspace{10mm}+t_{xy}c_{3j-\hat x-\hat y}+t_{xy}^*c_{3j+\hat x-\hat y})\big]
\nonumber\\
+&\mbox{H.c.} ,
\label{Hop_14}%---
\end{alignat}
where $t_j$ ($j=x,xy$) is a complex parameter which we set
$t_j=t_j^{\rm r}+it_j^{\rm i}$. It yields
\begin{alignat}1
h_{\rm 14}=&2\cos k_x(t_x^{\rm r} +t_{xy}^{\rm r}\cos k_y) \gamma_2
\nonumber\\
-&2\sin k_x(t_x^{\rm i}+t_{xy}^{\rm i}\cos k_y)i\gamma_1\gamma_5
\nonumber\\
+&2t_{xy}^{\rm r}\cos k_x\sin k_y \gamma_1+2t_{xy}^{\rm i}\sin k_x\sin k_y i\gamma_2\gamma_5.
\end{alignat}

Including all these terms, Hamiltonian in the Brillouin zone can be denoted as
\begin{alignat}1
\delta h(k)=&h_{\rm d}(k)+h_{12}(k)+h_{13}(k)+h_{14}(k)
\nonumber\\
=&R_{\rm ee}\cos k_x\cos k_y+R_{\rm oo}\sin k_x\sin k_y
\nonumber\\
&+R_{\rm eo}\cos k_x\sin k_y+R_{\rm oe}\sin k_x\cos k_y,
\end{alignat}
where $R_{ij}$ are matrices whose transformation properties under time reversal, particle-hole, and chiral symmetries are
summarized in Table \ref{t:gham}.
Generically, this additional Hamiltonian breaks $C_4$ symmetry: In the following case, this Hamiltonian recovers 
$C_4$ symmetry,
\begin{alignat}1
&u_x=u_y, \quad s^{\rm i}=0,
\nonumber\\
&v=v_{xy}=0, v_y=-v_x,
\nonumber\\
&t_y=t_x,\quad t_{yx}=t_{xy}^*.
\end{alignat}

%\bibliography{2d_bbh}

\end{document}